\documentclass{aa}  

\usepackage{graphicx,textcomp}
\usepackage[varg]{txfonts}
\usepackage[colorlinks = true, allcolors=blue]{hyperref}

\usepackage{natbib,twoopt}
\usepackage{orcidlink}
\usepackage{longtable}

\begin{document} 

\title{Open superclusters $\textrm{I}$:}
\subtitle{The most populated primordial groups of open clusters in the third quadrant of the Galactic disk}

\author{Juan Casado\orcidlink{0000-0003-4105-2520}\inst{1}
\and
Yasser Hendy\orcidlink{0000-0002-2356-8315}\inst{2}\fnmsep\thanks{yasserhendy@nriag.sci.eg}
      }     
\institute{
$^{1}$ Facultad de Ciencias, Universitat Autònoma de Barcelona (UAB), 08193, Bellaterra, Barcelona, Spain\\
        \email{juan.casado@uab.cat}\\ 
$^{2}$ Astronomy Department, National Research Institute of Astronomy and Geophysics (NRIAG), 11421, Helwan, Cairo, Egypt\\
        } 

\date{Received---- ; accepted----}

\abstract
{We define an open supercluster (OSC) as a cluster of at least six open clusters (OCs) born from the same
giant molecular cloud (GMC). We surveyed the recent catalogs of OCs based on Gaia data and relevant
literature to find 17 OSCs of the third Galactic quadrant, along with 190 likely members of them. 
OSCs are frequent enough to be considered an extra class of objects in the hierarchy of star formation. Some of these
supersystems are new and most of them contain more members than previously thought. The detailed study
of some OSCs lead to the discovery of four new young OCs that are members of them, named Casado-Hendy 2 to 5. 
In certain instances, subgroups with distinct proper motions (PMs) or 3D positions have been found within an
OSC, suggesting the presence of multiple generations of stars formed from several bursts of star formation
within the same GMC. OSCs are typically unbound and tend to disintegrate on timescales of $\sim$ 0.1 Gyr. The
present results confirm that young OCs tend to form primordial groups and suggest that globular clusters
(GCs) are not formed from the accretion of OSCs, that is, at least in the local Universe at late times.}
 
               
\maketitle
%
\section{Introduction} \label{sec_1}
OCs and most stars are formed as embedded clusters within the cores of giant molecular
clouds (GMCs) \citep{Lada2003ARA&A..41...57L,Elmegreen_2006astro.ph.10679E}. These clusters represent the stellar groupings
associated with the inner and densest regions of the gas and are the result of star formation in hierarchically
structured gas clouds.Studies have shown that embedded star clusters are rarely born alone \citep{Bica_2003A&A...397..177B,Roman-Zuniga_2015AJ....150...80R,Camargo_2016MNRAS.455.3126C,Hao_2023RAA....23g5023H}. Instead, evidence suggests that OCs are usually born as
part of ``families'' related to a particular star-formation complex, known as cluster complexes \citep{Piskunov_2006A&A...445..545P,Conrad_2017A&A...600A.106C} or primordial groups, which have been confirmed by Gaia data 
(e.g., \citealp{Kounkel_2020AJ....160..279K,Casado_2021ARep...65..755C,Casado_2023MNRAS.521.1399C}).
Thus, cluster complexes or primordial groups refer to
groupings of clusters that are spatially close to each other and share a common origin from the same GMC.
These primordial groups are habitually young, with ages ranging from a few million to several tens of millions of
years. They typically contain associated coronae stars and nebulae and can be found mainly in the spiral
arms (e.g., \citealp{Gusev_2013MNRAS.434..313G}). A significant fraction of young OCs within 3 kpc of the Sun
belong to groups some hundreds of parsecs across \citep{Efremov_1988SvAL...14..347E}, and similar cluster complexes
have been found in other galaxies (e.g., 30 Doradus in the LMC; \citealt{Walborn_2002AJ....124.1601W}). For example, \citet{Grasha_2015ApJ...815...93G} studied the galaxy NGC 628, and found that the clustering of young clusters decreases with
cluster age.\\

The study of cluster aggregates is important because they provide insights into the formation and evolution
of star clusters, as well as larger-scale processes that govern the evolution of galaxies. The properties of
individual clusters within a group allow us to learn about their formation conditions and interactions. The term supercluster usually refers to clusters of galaxies but has also been used to describe associations
of stars. A supercluster of stars was defined as a group of gravitationally unbound stars that share a common
motion in space and occupy extended regions in the Galaxy \citep{Eggen_1994gsso.conf..191E}. The term has also been used to refer to a cluster of OCs. In strong starbursts, star clusters form in superclusters containing dozens to hundreds of young massive OCs within regions several hundred parsecs across \citep{Kroupa_1998MNRAS.300..200K} and with masses in the range of $10^6 - 10^8$ solar masses \citep{Fellhauer_2005ApJ...630..879F}. These superclusters may merge to become objects similar to faint GCs or spheroidal dwarf galaxies.\\

In this article, we define an OSC as a populated primordial group of OCs that
share similar positions and PMs in space, regardless of whether or not they are gravitationally
bound. While primordial groups with only two members have been found, the average primordial group has three or four members \citep{Casado_2021ARep...65..755C}. Known groups with seven or more OCs have so far been too scarce
to obtain a statistically significant sample \citep{Conrad_2017A&A...600A.106C,Liu_2019ApJS..245...32L,Casado_2021ARep...65..755C}. Therefore, we selected six as the minimum number of members for an OSC.\\

\citet{Casado_2021ARep...65..755C} used Gaia EDR3 data to identify 22 candidate primordial groups containing a total of 80 OCs between Galactic longitudes of 240 and 270 degrees. Most of those primordial groups are formed by young OCs ($\leq$ 0.1 Gyr). \citet{Casado_2022Univ....8..113C} confirmed that young clusters are more likely to be close to their siblings than old clusters. In \citet{Casado_2023MNRAS.521.1399C}, we discussed the primordial group of NGC 6871 in detail and identified two new young OCs as members, concluding that all members originated from the same GMC, but that this group is rapidly disintegrating.\\

The present article is a continuation of our previous work on the topic. We identify, list, and describe OSCs in the
third quadrant of the Galactic disk, including their probable and possible candidate members. Section \ref{sec_2} 
describes the methods and criteria used to identify candidate OCs for the listed OSCs. Section \ref{sec_3}  summarizes the general landscape of OSCs in the solar neighborhood of the Galactic disk. The subsections of Section
\ref{sec_3} provide an analysis of each OSC found in the third quadrant of the Galaxy. Section \ref{sec_4}  includes a general discussion of our findings and the main conclusions of our study.
\section{Methods} \label{sec_2}

To select our sources, we used the Gaia-derived cluster catalogs of \citet[][2443 clusters]{Liu_2019ApJS..245...32L}, \citet[][2017 OCs]{Cantat-Gaudin_2020A&A...640A...1C}, \citet[][1743 clusters]{Dias_2021MNRAS.504..356D}, \citet[][3794 clusters]{Hao_2021A&A...652A.102H}, \citet[][1872 clusters]{Poggio_2021A&A...651A.104P} and \citet[][7167 clusters]{Hunt_2023A&A...673A.114H} --- which are the only all-sky catalogs that have thousands of OCs with all five astrometric Gaia measurements (position, PMs, and parallax) and ages ---, along with other literature sources. To select the related OCs, we used the five mean astrometric parameters of the listed objects to constrain the local overdensities, which can be observed in plots such as Fig. \ref{fig_1}. It should be emphasized that we scrutinized all the listed OCs --- not only the young ones --- for such overdensities using a method equivalent to that used for the manual search for OCs from the Gaia data of stars, which has been described elsewhere \citep{Casado_2021ARep...65..755C,Casado_2023MNRAS.521.1399C}. This is why we obtained members older than 100 Myr in a few cases (Table \ref{tab_1}).\\

To feed the main Table \ref{tab_1}, we used the data of the catalog or reference that lists the greatest number of OCs as
candidate members of each OSC. Extra candidates from other catalogs or references are listed in individual
tables for each OSC. We refined the raw list of candidate members by filtering out the outliers using the
following process: We applied maximum radii $\Delta R$ of 150 pc and maximum tangential velocity versus the mean $\Delta V_{tang}$ of 10 km/s
of all the member clusters using Eq. (\ref{eq_1}) and Eq. (\ref{eq_2}), respectively.

\begin{equation} 
\label{eq_1}
{\Delta R}={{\sqrt{((l-\overline{l})^2+(b-\overline{b})^2)} \ ({\pi} \ 1000/180 \ \overline{plx}))}}
,\end{equation}   
   
\begin{equation} 
\label{eq_2}
{\Delta V_{tang}}={{\sqrt{(( \mu^{\star}_{\alpha} -\overline{\mu^{\star}_{\alpha}})^2+(\mu_{\delta}-\overline{\mu_{\delta}})^2)}      \ (4.7404 / \overline{plx}))}}
.\end{equation}    

The $\overline{l}$, $\overline{b}$, $\overline{\mu^{\star}_{\alpha}}$, $\overline{\mu^{\star}_{\delta}}$, and $\overline{plx}$ are the means of Galactic longitude, Galactic latitude, proper motion in right ascension (RA), proper motion in declination (Dec), and parallax, respectively. All selected members in the OSCs have uncertainty intervals of less than $2\, \sigma$ from the means of the five dimensions $l$, $b$, $\mu^{\star}_{\alpha}$, $\mu_{\delta}$, and $plx$, which stand for the Galactic coordinates, PMs and parallax, respectively. We apply a sufficient number of iterations to obtain convergence with the same number of OCs from the last iteration.\\

The threshold of 300 pc in size is somewhat arbitrary, but has been chosen while taking into account the typical
size of the aggregates found in the literature \citep{Efremov_1988SvAL...14..347E,Piskunov_2006A&A...445..545P,Elmegreen_2014ApJ...787L..15E,Conrad_2017A&A...600A.106C}. The maximum span in PM (20 km/s) is chosen, because the typical difference between related OCs is < 10 km/s (e.g., \citealp{Conrad_2017A&A...600A.106C,Casado_2021ARep...65..755C}), and the difference in radial velocity (RV) between two OCs in the same complex is $\approx 20$ km/s \citep{Fellhauer_2005ApJ...630..879F}. The remaining OCs in the raw list that do not pass this filtration process are also mentioned but are considered less likely candidates.\\

We downloaded the selected data from Gaia DR3 for the newfound OCs Casado-Hendy 2, Casado-Hendy
3, Casado-Hendy 4, and Casado-Hendy 5 in this study. We removed the stars with RUWE > 1.4 and with
G > 18, as the $plx$ and PM errors increase exponentially with magnitude. We used the PYUPMASK
algorithm to obtain the likely members of each OC with a membership probability of larger than 0.8,
outperforming UPMASK \citep{Krone-Martins_2014A&A...561A..57K}, as shown in \citet{Pera_2021A&A...650A.109P}. 
We calculated the fundamental parameters (age, distance, and extinction) using the ASTECA algorithm
\citep{Perren_2015A&A...576A...6P}. We made use of the PARSEC v1.2S \citep{2012MNRAS.427..127B} theoretical
isochrones (obtained from the CMD website\footnote{http://stev.oapd.inaf.it/cgi-bin/cmd}). We downloaded the
data of $\mathcal{Z}$ range (0.0001 to 0.06) with a step of 0.01 and log(age) range (6.0 to 10.15) with a step of 0.01.
\section{An overview of Galactic OSCs} \label{sec_3}
\begin{figure*}[!ht]
    \includegraphics[width=1\textwidth]{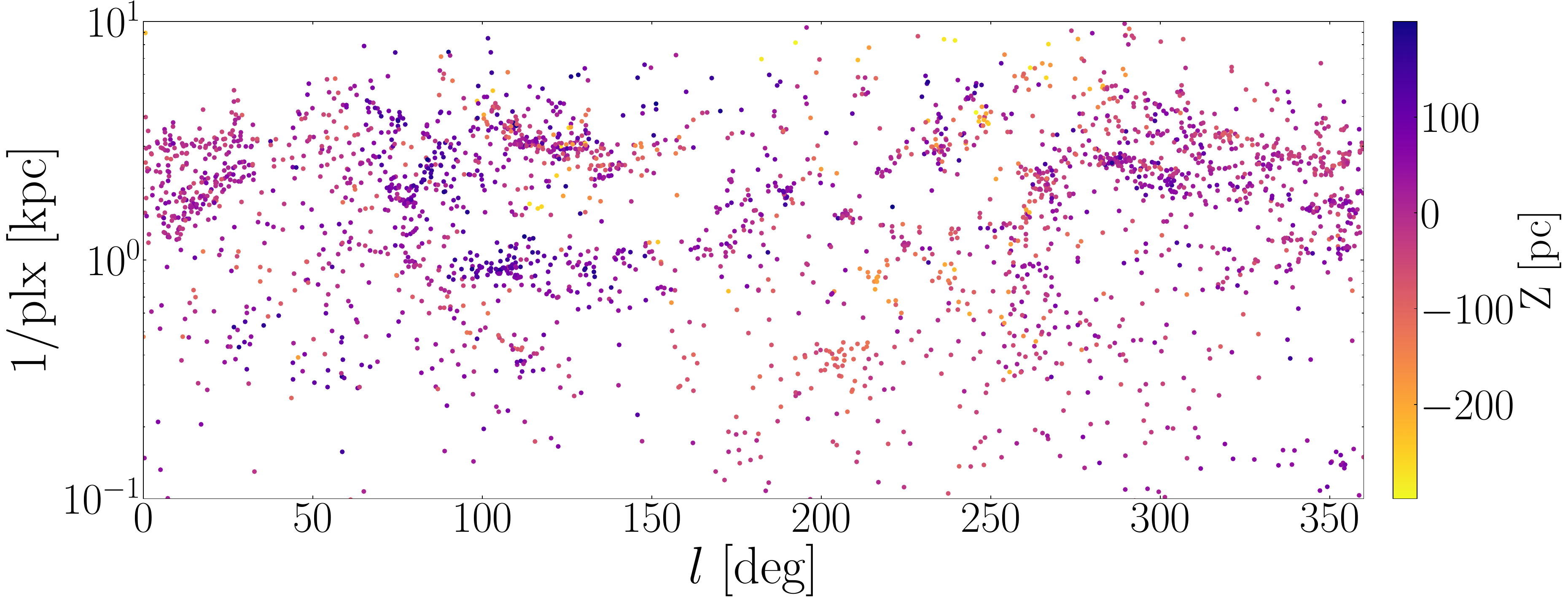}
    \caption{Galactic longitude--distance (1/plx) plot, colored according to altitude of the 2747 candidate clusters of < 100 Myr old from the \citet{Hunt_2023A&A...673A.114H} catalog. However, this latter work only focuses on the third quadrant of the Galaxy ($180^{\circ} < l < 270^{\circ}$).}
    \label{fig_1}
\end{figure*}

Figure \ref{fig_1} shows that clusters younger than 100 Myr form coherent superstructures. This reflects their
relatively recent formation in the same GMC. Similar colors in Fig. \ref{fig_1} indicate similar distances to the
Galactic plane (Z) and therefore a similar 3D position of the groups of OCs. Fig. \ref{fig_1} also shows prominent
elongated superstructures at long distances, which trace fragments of distant Galactic arms. For example,
the assembly at approximately 100 to 140 degrees longitude corresponds to the Perseus arm, while the
ensemble of OCs from approximately 0 to 30 degrees longitude can be attributed to the Sagittarius arm. On
the other hand, the oldest OCs do not show significant clustering \citep{Casado_2022Univ....8..113C}.\\

Additionally, we found a clump of at least six OCs (HSC 2907, HSC 2919, HSC 2931, OCSN 96,
OCSN 98, and OCSN 100) at an $l$ of approximately 355 degrees and a $plx$ of close to 7 mas, which is probably the closest
OSC to the Sun. This noteworthy system is part of the Upper Scorpius star forming region, but is beyond
the scope of this study. In the following subsections, we discuss the characteristics of the individual OSCs studied in this work and
their members. These OSCs are located in the third quadrant of the Galaxy and are ordered by increasing
longitude. Their main member candidates are summarized in Table \ref{tab_1}. Most OSCs in Table \ref{tab_1} can be
observed in Fig. \ref{fig_1}.
\subsection{Gem OB1} \label{sec_3-1}
This newly discovered OSC has at least seven well-grounded candidate members, shown in Table \ref{tab_1}. The
parameters of other likely candidates are summarized in Table \ref{tab_2}. The system corresponds to the OB
association Gem OB1, from which it derives its name. This star-forming system includes the HII region
NGC 2174 and the supernova remnant IC 443 \citep{Bica_2019AJ....157...12B}. A matching distance of 1.85 kpc has been
reported for Gem OB1 \citep{Zucker_2020A&A...633A..51Z}. From the positions of the most likely members and their median parallax, we derive a projected size of approximately 110 pc. From the distances of these members, we obtain a comparable radial size of 140 pc.
The maximum difference in tangential velocity is estimated to be 9 km/s based on PMs and median
parallax.\\

The mean RV of Pismis 27 (-8 km/s from 3 stars; \citealp{Hunt_2023A&A...673A.114H}) differs from that of other candidate
members, but it has a high standard deviation (38 km/s). Moreover, various RVs have been reported for
Pismis 27, including an RV of 28±10 km/s \citep{Tarricq_2021A&A...647A..19T}, which is similar to other siblings in the
OSC (Table \ref{tab_1}). Other potential members include IC 2157, FSR 902, Collinder 89, OC 315, Theia 3475, and the candidate
embedded cluster MWSC 0759 \citep{Casado_2022Univ....8..113C}.
\subsection{Ori OB1} \label{sec_3-2}
The Orion star-formation complex, part of the Gould Belt, contains a well-known primordial group, rich in
young OCs, that overlaps the Ori OB1 association, as well as several nebulae, such as M42 and M43 \citep{Bica_2019AJ....157...12B}. 
\citet{Chupina_2000ESASP.445..347C} reported that OCs Trapezium, NGC 1977, and NGC 1980, among other star groups, have almost parallel PMs. 
\citet{Elias_2009MNRAS.397....2E} detected 11 clusters associated with Ori OB1, namely NGC 1976, NGC 1977, NGC 1980,
NGC 1981, Collinder 70, Sigma Ori, ASCC 16, and ASCC 18-21. \citet{Conrad_2017A&A...600A.106C} studied the grouping
of OCs in the Galaxy using RVs and found only one complex containing 15 OCs, including Collinder 69,
Platais 6, ASCC 24, and NGC 2232 as new candidate members. Concerning NGC 2232, \citet{Pang_2020ApJ...900L...4P} 
related it to LP 2439. \citet{Liu_2019ApJS..245...32L} identified a group of seven OCs in this region based on Gaia 3D
position only. \citet{Piecka_2021A&A...649A..54P} stated that ASCC 19, Gulliver 6, UBC 17a, and UBC 17b form
an aggregate of OCs.\\

Our study found a total of 11 compatible clusters in the 5D astrometric space (Table \ref{tab_1}) using the catalog of
\citet{Cantat-Gaudin_2020A&A...640A...1C}, with 12 additional robust candidates detailed in Table \ref{tab_3}. We note that Gulliver
6 and UBC 17b appear to be the same OC due to their analogous parameters and shared member stars \citep{Hunt_2023A&A...673A.114H}. 
Our study also suggests that two of Liu \& Pang candidate members (LP 2368 and LP
2381) should be excluded from the group because of their disparate PMs. Other possible candidate members are UPK 385, UPK 398, UPK 422, LP 2384, LP 2439, OCSN 56, OCSN 70, CWNU 216, CWNU 1072, M78, NGC 2232, ASCC 24, Platais 6, and Collinder 69. Overall, this OSC
is one of the most populated (and closest) systems in the Galaxy.\\

\begin{figure}[!ht]
    \centering
    \includegraphics[width=0.51\textwidth]{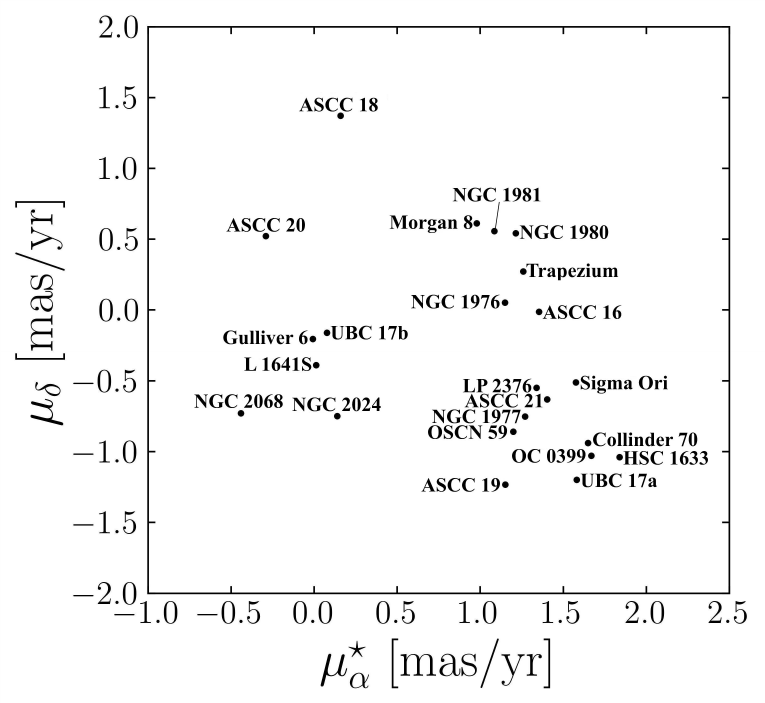}
    \caption{Vector position diagram  of the likely members of the Ori OB1 OSC. At least two moving subgroups are discerned.}
    \label{fig_2}
\end{figure}

\begin{figure}[!ht]
    \centering
    \includegraphics[width=0.51\textwidth]{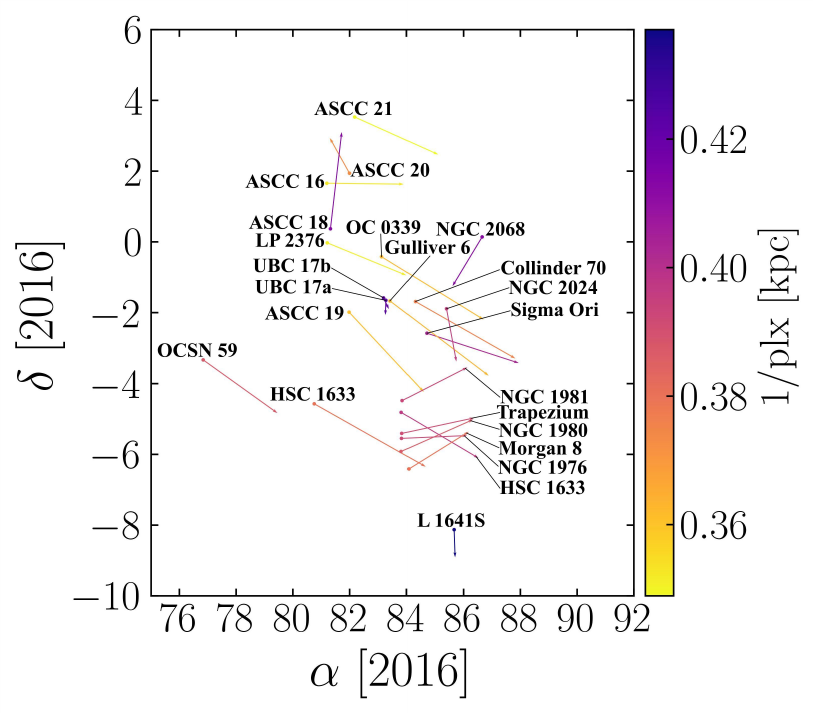}
    \caption{Positions and PMs of the OCs in Ori OB1. Vectors depart from the equatorial coordinates of each
member and are proportional to the mean PM.}
    \label{fig_3}
\end{figure}

The span in positions indicates that this OSC is approximately 90 pc wide, with a similar radial size obtained
from the range in parallaxes. The vector position diagram (VPD) of the likely members shows at least two
subgroups with different PMs and a few OCs with peculiar PMs (Fig. \ref{fig_2}). When PMs
are plotted as arrows from the position of each OC, Fig. \ref{fig_3} shows dispersion of members in diverse
directions. However, the most extreme PMs indicate a span in projected velocity of only 4 km/s.\\

Even adjacent, populated OCs NGC 1976 and Trapezium have somewhat divergent PMs. Considering the
two most separated OCs (L 1641S and ASCC 16), their PM vectors also indicate that they will become
more separated in the future. This is also true when considering other members instead of ASCC 16 (e.g.,
ASCC 18 or ASCC 20). These results suggest that this OSC is rather unbound and disintegrating.\\

According to \citet{Dias_2021MNRAS.504..356D}, all members except Sigma Ori (RV= -14 km/s) have RVs in the range of
15 to 32 km/s, indicating a RV dispersion of $\approx 17$ km/s. However, Sigma Ori has been regularly reported to
have an RV in the range of 28 to 31 km/s \citep{Dias_2002A&A...389..871D,Kharchenko_2013A&A...558A..53K,Conrad_2017A&A...600A.106C,Tarricq_2021A&A...647A..19T,Qin_2023ApJS..265...12Q},
similar to the rest of the members. The ten likely members in \citet{Hao_2021A&A...652A.102H}, 
with the exception of ASCC 19 (RV $\approx -1$ km/s), have RVs of between 17 and 33 km/s, leading to a very similar
RV distribution. RVs for ASCC 19 from 18 km/s \citep{Zhong_2020A&A...640A.127Z} to 24 km/s 
\citep{Dias_2002A&A...389..871D} have also been reported, compatible with the rest of the members. The vast majority of members found using
\citet{Hunt_2023A&A...673A.114H} data have RVs from 12 to 29 km/s, confirming a dispersion of $\approx 17$ km/s in RV.\\

All members of this OSC are young, with a maximum age range of 3 to 98 Myr combining all the catalogs
used. The oldest estimation corresponds to NGC 1977 in \citet{Cantat-Gaudin_2020A&A...640A...1C}. However, reported
ages of NGC 1977 from other authors range from 4 Myr \citep{Kharchenko_2013A&A...558A..53K} to 12 Myr \citep{Dias_2002A&A...389..871D}. 
The age of OC 0339 (Table \ref{tab_3}) is also uncertain, as an estimation of 6 Myr has been published by \citet{Hao_2022A&A...660A...4H}. The vast majority of members are up to 20 Myr old, and it is well known that star formation is currently in
progress in the region. Such a range of ages is enough to suggest diverse events of triggered star formation
within the same GMC.
\subsection{Mon OB2} \label{sec_3-3}
\citet{Conrad_2017A&A...600A.106C} proposed a pair of linked OCs: NGC 2244 and Collinder 107. 
\citet{Costado_2018MNRAS.476.3160C} related NGC 2244 and Collinder 106, while \citet{Piecka_2021A&A...649A..54P} confirmed that Collinder 106,
Collinder 107, and NGC 2244 are related. \citet{Muzic_2022A&A...668A..19M} noted that Collinder 106, Collinder 107, and
NGC 2237 have PMs and distances close to that of NGC 2244 and concluded that NGC 2244 shows a clear
expansion pattern and may be unbound.\\

According to our study, this OSC has nine likely members (Table \ref{tab_1}), and overlaps the OB association Mon
OB2, from which we take its name. The superstructure has an apparent projected size of approximately 70 pc and
a comparable radial size of approximately 110 pc, estimated from the positions of its members and their parallaxes and/or
photometric distances. The maximum difference in PMs implies a range in projected velocities of 8 km/s.
Reported RVs are too scattered to reach any robust conclusion (e.g., \citealp{Costado_2018MNRAS.476.3160C}), though most likely
members with a Gaia DR3 RV have values of around 30 km/s \citep{Hunt_2023A&A...673A.114H}. NGC 2237, considered
a satellite cluster of NGC 2244 in the Rosette nebula with a reported RV of 37.6 km/s \citep{Evans_1967IAUS...30...57E}, is
another likely member of this OSC. Dubious members of Mon OB2 OSC could include Coin-Gaia 28, UBC
83, UBC 620, UBC 1330, CWNU 2558, HSC 1657, Theia 1734, vdBergh 85, FSR 1085, and FSR 1144.
\subsection{Mon R2} \label{sec_3-4}
This OSC is related to Mon R2, a star association and HII region + GMC, from which we take its name. It contains
seven likely OCs (Table \ref{tab_1}), with parameters suggesting a projected size of approximately 120 pc and a
similar radial size. The RV of Mon R2 (36 km/s; \citealp{Kuhn_2019ApJ...870...32K}) agrees with five of the most likely members of the OSC.
We note that OC 0352 has an eccentric RV of 86$\pm$52 km/s (Table \ref{tab_1}), but a matching RV of 31$\pm$16 km/s has
also been reported by \citet{Hao_2022A&A...660A...4H}. Thus, the RV data corroborate the existence of this OSC.
Other possibly associated members could include UPK 416, UPK 436, UPK 448, Camargo 119, HSC 1732,
OC 0361, and vdBergh 80.
\subsection{Mon OB3} \label{sec_3-5}
This newly found OSC contains at least six likely members (Table \ref{tab_1}) and matches the association Mon
OB3, which contains the HII region Sh2-287. From the positions of the likely members and the median
parallax, a projected size of approximately 70 pc is estimated. The radial size derived the most likely distances
of the members leads to a similar value of 80 pc. The most diverse PMs among the likely members lead to
a tangential velocity dispersion of 9 km/s. The known RVs are rather spread but are compatible with a mean
value of around 40 km/s. Parameters of other candidate members are summarized in Table \ref{tab_4}. Less likely candidates would include
CWNU 1410 and 1447, FSR 1125 and 1149, HXHWL 74, OC 0369 \citep{Hunt_2023A&A...673A.114H}, and OC 74 in \citet{He_2021RAA....21...93H}.
\subsection{CMa OB1} \label{sec_3-6}
This system is related to the star-forming regions CMa OB1 and CMa R1 \citep{Gregorio-Hetem_2009A&A...506..711G}.
CMa R1 is associated with the bright-rimmed nebula Sh2-296, along with several embedded clusters and
B stars. \citet{Santos-Silva_2021MNRAS.508.1033S} performed a detailed study of the stellar groups in the CMa OB1 region.
However, their work encompasses 15 clusters with distances from 550 to 1560 pc and assorted PMs, making
it unclear which belong to this star-forming region, except for four new cluster candidates (CMa05 to
CMa08), which are indeed the youngest ones (< 20 Myr).\\

\begin{figure}[!ht]
    \centering
    \includegraphics[width=0.25\textwidth]{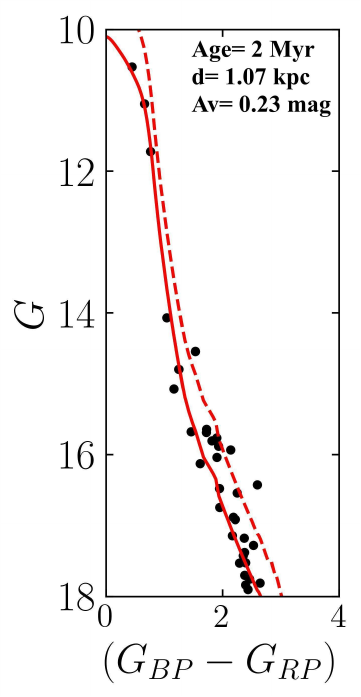}
    \caption{Color--magnitude diagram and the fitted isochrone of the new OC Casado-Hendy 4. The continuous lines are the PARSEC isochrone fitted to our data, while dashed ones represent the same isochrone but vertically shifted by -0.75 mag (the locus of unresolved binaries of equal-mass components).}
    \label{fig_4}
\end{figure}

\begin{figure}[!ht]
    \centering
    \includegraphics[width=0.25\textwidth]{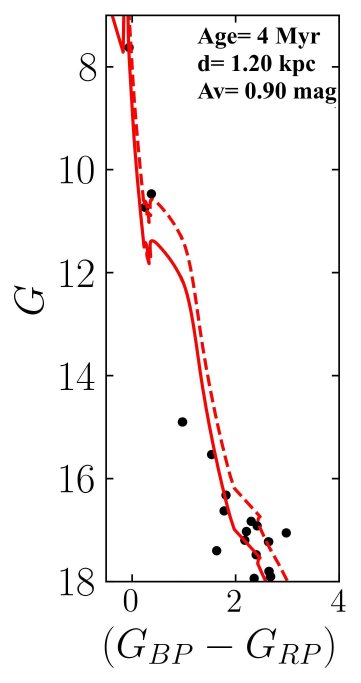}
    \caption{Color--magnitude diagram and the fitted isochrone of the new OC Casado-Hendy 5. The continuous lines are the PARSEC isochrone fitted to our data, while dashed ones represent the same isochrone but vertically shifted by -0.75 mag (the locus of unresolved binaries of equal-mass components).}
    \label{fig_5}
\end{figure}

In a first attempt, we found five member OCs from the \citet{Hunt_2023A&A...673A.114H} catalog (Table \ref{tab_1}), two of
which correspond to CMa06 and CMa08. A closer inspection of the region reveals two new member
clusters, the parameters of which are summarized in Table \ref{tab_5}. The ages and photometric distances in Table \ref{tab_5} were
estimated using the brightest (G<18) stars of Casado-Hendy 4 and Casado-Hendy 5 (Fig. \ref{fig_4} and \ref{fig_5},
respectively). The distance of Casado-Hendy 4 from both the inverted parallax and the Bayesian statistics
from stars in \citet{Bailer-Jones_2021AJ....161..147B} is 1.14 kpc. This new OC would have an estimated extinction of $A_V$ =0.23$\pm$0.13.
For Casado-Hendy 5, the estimated age and photometric distance are similar to its candidate siblings. The
parallax and Bayesian distances agree on a value of 1.18 kpc. The derived extinction is $A_V$ = 0.90$\pm$0.09. 
These results suggest that both OCs are very young and are robust candidate
members of the CMa OB1 supercluster.\\

\begin{figure}[!ht]
    \centering
    \includegraphics[width=0.48\textwidth]{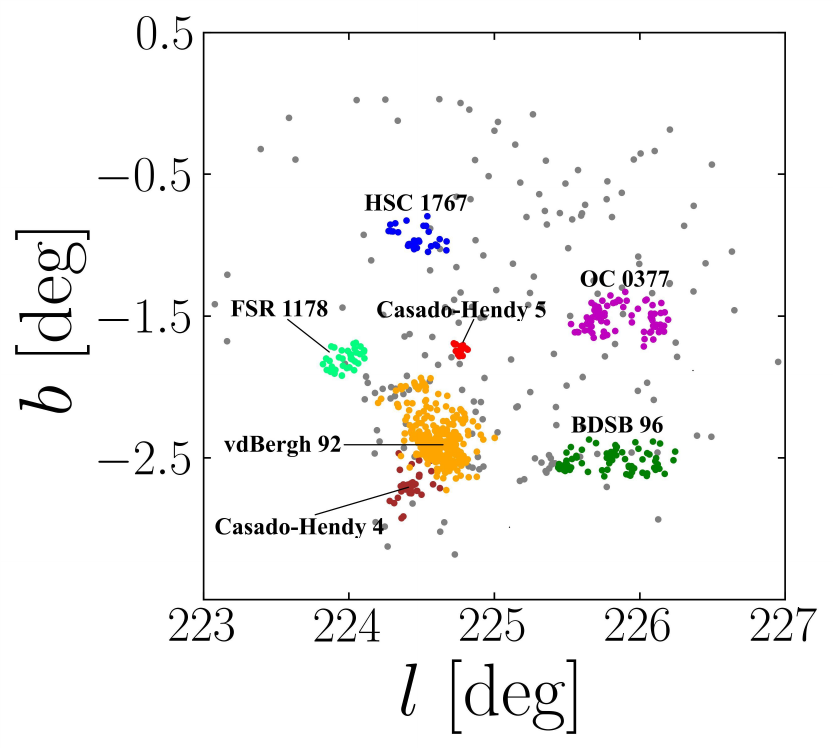}
    \caption{Member stars (G < 18) of the most likely OCs of the OSC CMa OB1, and some related but unaffiliated coronae stars (gray dots). The constraints are given in Table \ref{tab_1}.}
    \label{fig_6}
\end{figure}

From the parameters of the likely members, a projected size of approximately 35 pc is derived, while the
collection of parallaxes and distances point to a similar radial size of 40 pc. The PMs indicate a tangential
velocity width of 8 km/s, similar to the 6.1 km/s reported by \citet{Santos-Silva_2021MNRAS.508.1033S}, while the whole set
of RVs leads to a maximum RV dispersion of 17 km/s. The salient members of this OSC are plotted in Fig. \ref{fig_6}. It is apparent from this plot that some clusters
involve diverse clumps of stars (subclusters) and appear to be in the process of disintegration.\\

The OC vdBergh 92 is the main member of the system and coincides with the CMa R1 star-forming region.
Historically, vdBergh 92 had been considered to be at a distance of 1.5 kpc. However, the Gaia results
locate it between 1.11 kpc \citep{Hunt_2023A&A...673A.114H} and 1.19 kpc \citep{Cantat-Gaudin_2020A&A...640A...1C}. The median of
the reported parallaxes is 0.863 mas \citep{Poggio_2021A&A...651A.104P}, which corresponds to a distance of 1.16 kpc,
which is within the previous range and similar to the distances of Casado-Hendy 4. Casado-Hendy 4
seems to be at least 20 pc closer than vdBergh 92, but in the hypothetical case where both OCs are at the same
distance, they can form a binary system, as the distance among them would be less than 7 pc.\\

Five of our revisited star members of vdBergh 92 have a RV of 22$\pm$2 km/s, matching the result in \citet{Hunt_2023A&A...673A.114H}. 
Both measurements are within the general range of the members in Table \ref{tab_1}. One of the stars
of Casado-Hendy 4 has a RV from Gaia DR3 (Table \ref{tab_5}) that is also in agreement with the mentioned range. Other clusters likely associated with this OSC include CMa05, CMa07, and the embedded clusters in the
reflection nebulae NGC 2327 and BRC 27 \citep{Santos-Silva_2021MNRAS.508.1033S}. Accordingly, the distances of NGC
2327 and BRC 27 were reported as 1.2 kpc by \citet{Soares_2002A&A...388..172S}.

\subsection{HC2} \label{sec_3-7}
This newly described OSC has at least six likely members (Table \ref{tab_1}). The oldest member, HSC 1893,
appears to be older than usual members assembled in Table \ref{tab_1}. However, the confidence interval range (16
to 84\% percentile) for this unique reported age in the literature is wide: 90 to 250 Myr. The projected size of this system is 0.20 kpc, which is comparable to the radial size of 0.29 kpc. The span
of tangential velocities is 7 km/s. Reported RVs are too scarce to be of any help in defining this system. Other dubious members would include HSC 1841, UBC 1373 and 1378, and Theia 2110 \citep{Hunt_2023A&A...673A.114H}.
This OSC has comparable parameters and sizes to the following OSC HC3, suggesting that both
systems might be related.
\subsection{HC3} \label{sec_3-8}
\begin{figure}[!ht]
    \centering
    \includegraphics[width=0.25\textwidth]{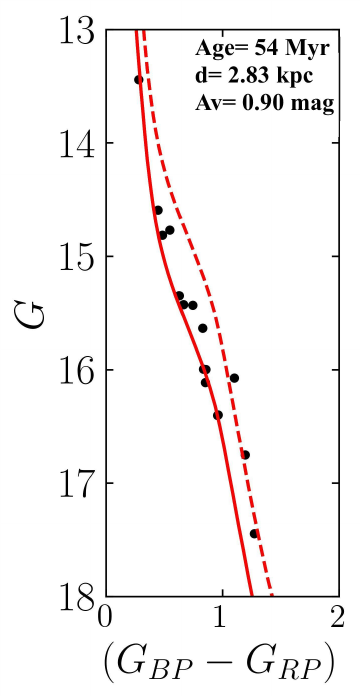}
    \caption{Color--magnitude diagram and the fitted isochrone of the new OC Casado-Hendy 2, likely member of the OSC HC3. The continuous lines are the PARSEC isochrone fitted to our data, while dashed lines represent the same isochrone but vertically shifted by -0.75 mag (the locus of unresolved binaries of equal-mass components).}
    \label{fig_7}
\end{figure}

The pair formed by NGC 2383 and NGC 2384 has been ruled out as a candidate binary system \citep{Angelo_2022MNRAS.510.5695A}.
However, we found eight OCs that form a primordial group including NGC 2384 (Table \ref{tab_1}).
All astrometric parameters (and ages) seem compatible. We also discovered a new OC in the studied field, Casado-Hendy 2, with parameters detailed in Table \ref{tab_6}. The isochrone fitting of this new member is given in Fig.\ref{fig_7}. From it, an age in the range of 50 to 58
Myr is obtained, compatible with the rest of the OSC members (Table A.1). The cluster extinction is 0.90$\pm$
0.06 mag, and the photometric distance is 2.83$\pm$0.06 kpc. The 16 likely stars in this OC are recorded in the
\citet{Bailer-Jones_2021AJ....161..147B} catalog. The median geometric distance to them is 2.87 kpc. From the corrected
median parallax of the cluster, a distance of 2.97 kpc is derived. Thus, the three distances are rather
consistent and it is quite safe to state that Casado-Hendy 2 is at a heliocentric distance of 2.9 kpc, within
the range of its siblings.\\

The projected size of the whole system is 0.20 kpc, and the radial size from the parallaxes of the members
(Table \ref{tab_1}) is similar: 0.25 kpc. Regarding the tangential velocities, the maximum difference among members
is only 5 km/s. Other candidate members of this OSC may be Czernik 31, UBC 460, UBC1376, and PHOC 16 (=LP 733)
\citep{Hunt_2023A&A...673A.114H}. The range of RVs is wide (Table \ref{tab_1}). However, the RVs of NGC 2384 (46$\pm$4 km/s;
\citealp{Conrad_2017A&A...600A.106C}), HSC 1875 (59 km/s; \citealp{Hunt_2023A&A...673A.114H}), and LP 733 (56 km/s; \citealp{Dias_2021MNRAS.504..356D}) are
close enough to allow at least these three OCs to be related.\\

\begin{figure}[!ht]
    \centering
    \includegraphics[width=0.48\textwidth]{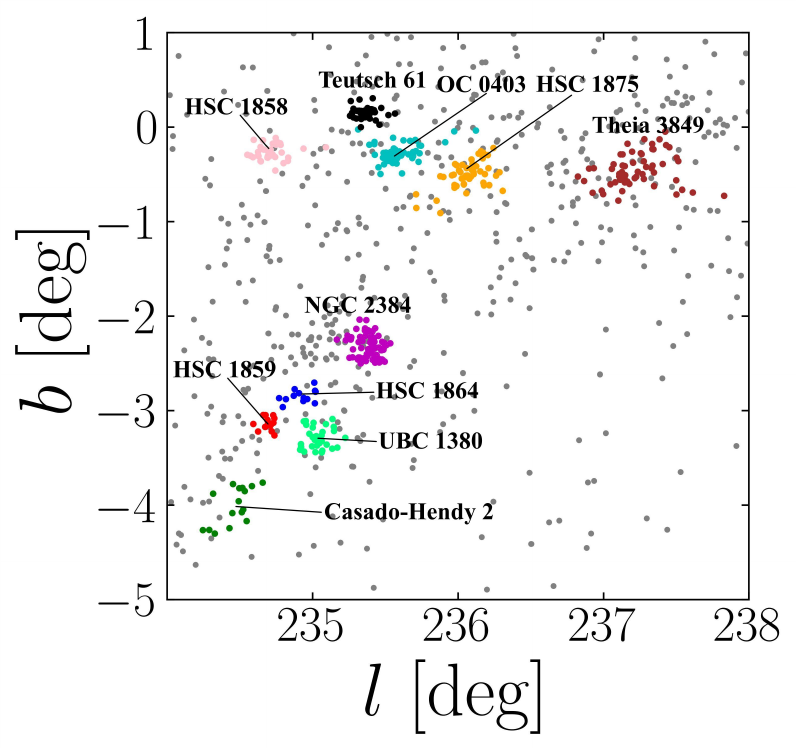}
    \caption{Chart of the member stars of the OSC HC3 down to G=18. The constraints are given in Table \ref{tab_1}.}
    \label{fig_8}
\end{figure}

Additional member candidates would be DBSB 7 and Ruprecht 16 \citep{Dias_2021MNRAS.504..356D}, UBC 224 (\citealp{Cantat-Gaudin_2020A&A...640A...1C}), Mayer 3, BDSB 6, Bochum 6, and LP198 (\citealp{Hao_2021A&A...652A.102H}). However, these are dubious members as they are not apparent in the chart of member stars (Fig. \ref{fig_8}). Figure \ref{fig_8} suggests two distinct subgroups of clusters forming two strings: One upper horizontal string, and a
second one containing NGC 2384. An overdensity of associated stars just left of NGC 2384 suggests that its
size and number of stars could be underestimated and confirms that this young OC is dissolving \citep{Angelo_2022MNRAS.510.5695A}.

\subsection{OSC of NGC 2362} \label{sec_3-9}
\citet{Liu_2019ApJS..245...32L} proposed a candidate pair of OCs including NGC 2354 and NGC 2362 based on 3D
position only. Their PMs are also compatible. \citet{Song_2022A&A...666A..75S} proposed a binary cluster involving
NGC 2362 and FSR 1297. We identified six firm candidate members for this OSC (Table \ref{tab_1}), including the cited NGC 2354,
NGC 2362, and FSR 1297(= Majaess 88). NGC 2362 is associated to its own GMC at a heliocentric distance
of 1.32$\pm$0.07 kpc (\citealp{Zucker_2020A&A...633A..51Z}). 
This distance overlaps the photometric and parallax-derived distance
ranges from the OSC members.\\

The projected size of the system is around 70 pc, while the radial size from the $plx$ and most likely distances
reported by \citet{Hunt_2023A&A...673A.114H} converge at a value near 120 pc. Regarding the tangential velocities,
there is an acceptable range among the most divergent members of 16 km/s. Exceptionally, a good account
of the RV of the likely members is available (Table \ref{tab_1}). The range in RV is 10 km/s, which is comparable to the range
of tangential velocities. These results support that the cited members genuinely belong to the same OSC.\\

However, the age of NGC 2354 appears to be too old for an OC in a primordial group. A review of the
literature reveals conflicting reports about the age of this OC. Some studies, such as those by 
\citet{Dias_2002A&A...389..871D}, \citet{Liu_2019ApJS..245...32L}, \citet{Cantat-Gaudin_2020A&A...640A...1C}, \citet{Tarricq_2021A&A...647A..19T}, \citet{Dias_2021MNRAS.504..356D}, and \citet{Hunt_2023A&A...673A.114H}, report a log(age) of approximately 9.1. 
In contrast, other authors, including 
\citet{Chen_2003AJ....125.1397C}, \citet{van-den-Bergh_2006AJ....131.1559V}, \citet{Wu_2009MNRAS.399.2146W}, \citet{Vande-Putte_2010MNRAS.407.2109V}, 
\citet{Gozha_2012AstL...38..506G}, \citet{Conrad_2017A&A...600A.106C}, and \citet{Loktin_2017AstBu..72..257L},
suggest a much younger age with a log(age) of around 8.1. \citet{Kharchenko_2013A&A...558A..53K} provide an estimate that falls between these two groups, with a log(age) of 8.61. The CMD of NGC 2354 displays a prominent red clump, blue stragglers, and other characteristics typical
of an old cluster. Our calculations yield a log(age) of 9.11$\pm$0.03. Studies suggesting a younger age for the
cluster are based on pre-Gaia data, while all studies using Gaia data concur on an older age. Therefore, we
believe that this OC is indeed relatively old. This result suggests that NGC 2354 is unlikely to be a member of this OSC. Several of its average
fundamental parameters, such as photometric distance (1.17 kpc), $\mu_{\delta}$ (1.8 mas/yr), and RV (32 km/s), lie at
the edges of the group’s range (see Table \ref{tab_1}). Furthermore, the CMDs of NGC 2354 and NGC 2362 do
not coincide as would be expected if they were to originate from the same GMC. Extra candidate members of this OSC may also be HSC 1905, Majaess 90, Theia 2000, and Theia 3099 according to data in \citet{Hunt_2023A&A...673A.114H}.
\subsection{HC4} \label{sec_3-10}
\citet{Casado_2021ARep...65..755C} proposed a new primordial group with six candidate members, including NGC 2453, FSR
1315, and four newly discovered OCs (Table \ref{tab_1}). In particular, NGC 2453 and FSR 1315 have compatible
Gaia astrometry, photometric distances, RVs, and ages. The parameters of other candidate members of this OSC are listed in Table \ref{tab_7}. UBC 465 presents a double
core. The secondary core coincides with Waterloo 3 \citep{Casado_2021ARep...65..755C}, which has very similar physical
characteristics. We updated the parameters of Casado 57 \citep{Casado_2021ARep...65..755C} using Gaia DR3 data. In
addition to the results shown in Table \ref{tab_7}, we note that Casado 57 has a consistent distance, disregarding the
method used to obtain it. Its estimated extinction is $A_V$ = 0.64$\pm$0.05 mag.\\

Additional, less likely candidates could be Haffner 19, NGC 2467, UBC 639, Ruprecht 32, Ruprecht 44,
and HSC 1952 (\citealp{Cantat-Gaudin_2020A&A...640A...1C}; \citealp{Casado_2021ARep...65..755C}; \citealp{Hunt_2023A&A...673A.114H}). We found reported RVs of
member candidates near 60 km/s --- with the exception of UBC 639 (93 km/s; \citealp{Casado_2021ARep...65..755C}): from 55.8$\pm$9.7 km/s for NGC
2467 (\citealp{Conrad_2017A&A...600A.106C}) to 64.7$\pm$0.4 km/s for NGC 2453 \citep{Dias_2019MNRAS.486.5726D}. As expected for a primordial
group, at least all the robust candidates are still embedded clusters or young OCs with reported ages of < 100
Myr.\\

Two subgroups with slightly different $\mu_{\delta}$ can be discerned within this OSC (Table \ref{tab_7}). However, the spread
in PM of all the likely members is < 20 km/s, while the apparent projected size is around 0.25 kpc. In any case,
we should be cautious regarding this candidate system because the cited members are around 5 kpc away,
near the precision limit of Gaia astrometric measurements. Thus, the high relative errors in $plx$ and distances
render any calculation of the OSC radial size  useless.
\subsection{Vel OB2} \label{sec_3-11}
According to \citet{Piskunov_2006A&A...445..545P}, this system is part of an extended open cluster complex associated
with the Gould Belt in the Vela-Puppis star formation region, perhaps the best example of clustered star
formation in GMCs. The OSC matches the Vel OB2 star association, which contains the Gum nebula \citep{Bica_2019AJ....157...12B}. 
\citet{Caballero_2008AN....329..801C} identified some groups of early stars (Escorial 24-30) in Vel OB2.
These groups contain 6 to 21 stars and, in some cases, belong to known OCs such as NGC 2451B, NGC
2547, or vdBH 23. \citet{Conrad_2017A&A...600A.106C} proposed a group of OCs containing vdBH 23, Trumpler 10, and
ASCC 48, but ascribed other candidate members to different groups.\\

Using Gaia DR1 data, \citet{Damiani_2017A&A...602L...1D} showed the presence of two dynamically distinct populations
in the area; one of them involves Vel OB2, while the other includes NGC 2547. Collinder 135, UBC 7,
Collinder 140, BBJ 1, NGC 2451B, and NGC 2547 have also been proposed as members of this system
based on Gaia data \citep{Soubiran_2018A&A...619A.155S,Beccari_2020MNRAS.491.2205B}. 
\citet{Cantat-Gaudin_2019A&A...621A.115C} identified seven
OCs in the same complex, namely Trumpler 10, vdBH 23, NGC 2451B, NGC 2547, Collinder 135,
Collinder 140, and UBC 7. \citet{Liu_2019ApJS..245...32L} identified a large group of 15 OCs. 
\citet{Kovaleva_2020A&A...642L...4K} established that Collinder 135 and UBC 7 form a binary cluster. 
\citet{Wang_2022MNRAS.513..503W} identified 13
OCs forming a snake-like structure: NGC 2232, Tian 2, Trumpler 10, Haffner 13, NGC 2451B,
NGC 2547, Collinder 132, Collinder 135, Collinder 140, UBC 7, BBJ 1, BBJ 2, and BBJ 3. The ``snake''
spreads over 400 pc and can be split into two main strings, encompassing the Vel OB2 OSC, and stretching
to the Orion constellation. The whole structure is most likely part of the Gould Belt, but disentangling which
members belong to which OSC is difficult, as it contains OCs associated to Vel OB2 and OCs related
to Ori OB1.\\

We revisited this interesting system and found 19 members in \citet{Hunt_2023A&A...673A.114H}, which are
summarized in Table \ref{tab_1}. Other likely members are detailed in Table \ref{tab_8}. However, two of the candidate
members of the cited Liu \& Pang group (NGC 2451A and IC 2391) appear to belong to the Sco-Cen OSC
according to our study. The likely members have a projected size of 0.19 kpc, which is comparable to a radial size of 0.14 to 0.15 kpc
depending of the distance considered. Regarding the PMs, the spread in projected velocity is about 15 km/s.
There are two outliers in RV (Table \ref{tab_1}), but 17 likely members have RVs from 8 to 22 km/s, leading to a
similar width of 14 km/s.\\

Regarding the ages, all the likely members are relatively young, but the age distribution strongly suggests
star formation in at least two steps, and extending to the present epoch. \citet{Cantat-Gaudin_2019A&A...621A.115C}
studied this system extensively and depict a scenario with several bursts of star formation too. Uncertain candidate members would be Escorial 25, Collinder 121, Collinder 132, Collinder 173, Pismis
4, LP 2385, LP 2388, Gulliver 9, Alessi 3, BBJ 1, BBJ 2, BBJ 3, Haffner 13, Theia 249, NGC 2516, ASCC
44, and ASCC 48 \citep{Conrad_2017A&A...600A.106C,Caballero_2008AN....329..801C,Bica_2019AJ....157...12B,Liu_2019ApJS..245...32L,
Beccari_2020MNRAS.491.2205B,Cantat-Gaudin_2020A&A...640A...1C,Wang_2022MNRAS.513..503W,Hunt_2023A&A...673A.114H}. 
Thus, the whole structure could encompass up to 39 members. However, ASCC 48 is a doubtful case as it has not appeared in any
Gaia study.
\subsection{HC5} \label{sec_3-12}
\citet{Casado_2021ARep...65..755C} proposed a candidate primordial group with four substantiated members (FSR 1342, LP
204, Ruprecht 47 and 48) and two doubtful ones (MWSC 1420 and Ruprecht 54). The revised HC5 OSC
using the \citet{Dias_2021MNRAS.504..356D} catalog contains four extra candidates (Table \ref{tab_1}). The OSC has a projected size of 0.20 kpc and a similar radial size: between 0.14 kpc (from $plx$) and 0.21
kpc (from photometric distance) according to data in \citet{Hunt_2023A&A...673A.114H}. From the most extreme PMs,
a span in tangential velocity of 8 km/s is estimated, while the span in RV is around 15 km/s (Table \ref{tab_1}).\\

The age estimated by \citet{Dias_2021MNRAS.504..356D} for FSR 1342 (0.26 Gyr) is higher than expected for a primordial
group, but an age of only 0.004 Gyr has also been reported \citep{Liu_2019ApJS..245...32L}. In any case, it should be noted
that several of the members of this OSC seem to be more than 100 Myr old. For instance, five likely
members in \citet{Hunt_2023A&A...673A.114H} have ages in the range from 0.04 Gyr (FSR 1342) to 0.25 Gyr (UBC
1420). The relatively high ages make this system the oldest OSC of the present work.\\

Three other likely candidates are detailed in Table \ref{tab_9}, which gives a total of 11 bona fide members. Several
of them seem to be OCs with more than 100 stars. For instance, Ruprecht 48 has 342 member stars \citep{Hunt_2023A&A...673A.114H},
whilst LP 204 and LP 942 have more than 600 members each \citep{Dias_2021MNRAS.504..356D}. 
These high populations help to explain the above-average stability of this OSC. Other possible candidates would
be Casado 9, Casado 10, LP 1429, LP 1454, and Ruprecht 55. If some of these candidates were authentic
members, the relative stability of the whole system would also be favored.
\subsection{HC6} \label{sec_3-13}
\citet{Casado_2021ARep...65..755C} postulated a primordial group candidate with eight members, the constraints on which are
summarized in Table \ref{tab_1}. This OSC is dominated by the highly populated cluster NGC 2439 (612 stars). A detailed inspection of the \citet{Cantat-Gaudin_2020A&A...640A...1C} and \citet{Hunt_2023A&A...673A.114H} 
catalogs reveals three other likely candidates (Table \ref{tab_10}). Other possible candidates from these catalogs are ESO 368-14, UBC 468,
HSC 1979, and HSC 2001.\\

The projected size is 0.21 kpc, whilst the radial size from the corrected parallaxes is similar: 0.23 kpc. The
span in tangential velocities is near 7 km/s. Reported RVs of the likely members vary relatively widely (Table
\ref{tab_1}), but the RVs in \citet{Hunt_2023A&A...673A.114H} are more consistent: from 52 km/s (Haffner 15) to 68 km/s (NGC
2439), leading to an acceptable dispersion of 16 km/s. The ascription of Bochum 15 to the group   by
\citet{Casado_2021ARep...65..755C} is
uncertain, but new data from \citet{Hunt_2023A&A...673A.114H}, including RV (64 km/s) and the
most likely distance (3.4 kpc), closely match the rest of the siblings, thus confirming its membership. The comparable characteristics of HC6 and HC5 OSCs suggests that both systems could be related.
\subsection{Sco-Cen} \label{sec_3-14}
The Sco-Cen association includes Sco OB2 and Upper Centaurus-Lupus \citep{Bica_2019AJ....157...12B}. 
Although \citet{Elias_2009MNRAS.397....2E} examined this OB association, these authors did not find any related cluster. 
\citet{Conrad_2017A&A...600A.106C} identified three likely members: NGC 2451A, IC 2391, and Platais 9. \citet{Eggen_1991AJ....102.2028E} discovered a stream of several dozen stars associated with IC 2391 and used the term ``supercluster'' to describe it. A possible genetic
relationship between IC 2391 and the associated stream has been detected \citep{Postnikova_2020RAA....20...16P}. The cluster and stream appear to have formed in the same region of the Galactic disk even though their present locations differ somewhat.\\

We found six likely members, the constraints of which are summarized in Table \ref{tab_1}. However, two of our
likely members, Platais 8 and IC 2602, were ascribed to a different group by \citet{Conrad_2017A&A...600A.106C}. 
Other possible member candidates would be HSC 2026, HSC 2068, HSC 2139, HSC 2237, and HSC 2351
\citep{Hunt_2023A&A...673A.114H}. Sco-Cen appears to be very extended OSC in the sky ($37^{\circ}$), but its real size is regular. From the coordinates
of the most distant members, the projected size is about 100 pc. The radial size is about 60 pc. Regarding
the projected velocities, the span in PMs leads to a spread of 11 km/s, while the spread in RV is very similar, of
12 km/s (Table \ref{tab_1}), supporting the idea of a coherent OSC. All candidate members resulting from 5D
astrometric selection are also young and have consistent ages (Table \ref{tab_1}). Altogether, Sco-Cen is a firm candidate
to be an authentic OSC. This OSC is also part of the Gould Belt, and is therefore related to Vel OB2 and Ori OB1.
\subsection{Vela Molecular Ridge} \label{sec_3-15}
\citet{Liu_2019ApJS..245...32L} proposed a triplet of OCs comprising vdBH 56, 
Collinder 197, and Alessi 43 in this region. \citet{Casado_2021ARep...65..755C} reported a primordial group candidate containing four members: Pismis 5, Ruprecht 64, Collinder 197, and MWSC 1579 (=Alessi 43). All of them are young clusters. Subsequently, 
\citet{Qin_2023ApJS..265...12Q} proposed a binary cluster candidate (Alessi 43 and Collinder 197) and a triple cluster candidate
(Pismis 5, Pismis 5A, and Pismis 5B). Authors suggested that these five clusters are a large cluster
aggregate. They also studied three possibly related OCs: vdBH 56, NGC 2546, and the new cluster candidate
QZ1. However, the parameters of this latter  agree with those of Ruprech 64 \citep{2021A&A...646A.104H}, to the point that QZ1 and Ruprech 64 are likely the same object.\\

In the present study, we found six members of this OSC using the \citet{Liu_2019ApJS..245...32L} 
catalog (Table \ref{tab_1}). 
The parameters of another likely member from \citet{Cantat-Gaudin_2020A&A...640A...1C}, Pismis 5, and a new OC
discovered in the present work, Casado-Hendy 3, are detailed in Table \ref{tab_11}. Thus, there is a total of eight likely
members, most of them consistent with the literature. Another, dubious candidate would be UPK 528. The
\citet{Dias_2021MNRAS.504..356D} catalog leads to a reliable group containing Alessi 43, vdBH 56, Collinder 197,
Ruprecht 64, and Pismis 5. From their PMs, a maximum velocity difference between the members of around
10 km/s is estimated, while the RVs span reasonably from 20 to 41 km/s.\\

\begin{figure}[!ht]
    \centering
    \includegraphics[width=0.25\textwidth]{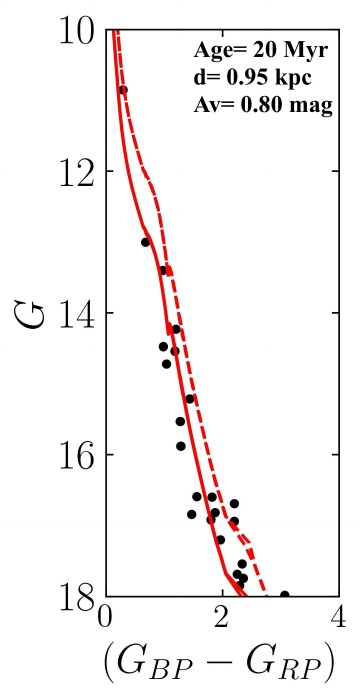}
    \caption{Color--magnitude diagram and the fitted isochrone of the new OC Casado-Hendy 3. The continuous lines are the PARSEC isochrone fitted to our data, while dashed lines represent the same isochrone but vertically shifted by -0.75 mag (the locus of unresolved binaries of equal-mass components).}
    \label{fig_9}
\end{figure}

The Galactic longitude ($l$ = $267^\circ$) of Casado-Hendy 3 is slightly different from that of the other members (from
$254^\circ$ to $265^\circ$), but the rest of the astrometric measurements fit well with the general parameters of this OSC
(Table \ref{tab_1}). Moreover, the mean RV of the new OC (20 km/s) --- obtained from four likely member stars ---
matches the mentioned range of RVs of the other OSC members. Furthermore, the age of this new
cluster (19 to 22 Myr) --- obtained by isochrone fitting of the CMD (Fig. \ref{fig_9}) --- lies within the range of the other OSC members. The derived extinction is 0.80$\pm$0.06 mag. The distance from the median $plx$
(0.95$\pm$0.06 kpc) and the photometric distance (Table \ref{tab_11}) match perfectly, and also agree with the
Bayesian distance of the star members from \citet{Bailer-Jones_2021AJ....161..147B}: 0.93 kpc.\\

Considering the whole OSC, the projected size is around 170 pc. Using photometric distances from 
\citet{Cantat-Gaudin_2020A&A...640A...1C}, the group seems to span from 0.84 to 0.96 kpc, a range of around 120 pc, a comparable radial
size. All the likely members are less than 100 Myr old. All considered, the Vela Molecular Ridge appears to be a robust OSC candidate.

\subsection{HC7} \label{sec_3-16}
This newly discovered OSC contains at least six likely members (Table \ref{tab_1}) and does not seem to be linked
to any known OB association or active star-forming region. The projected size of HC7 is approximately 60 pc, which is comparable to the radial size from the data in Table \ref{tab_1} ($\approx$ 100 pc).
Concerning the tangential velocities, we estimate a dispersion of 10 km/s. The RVs of the likely members
have a higher dispersion (Table \ref{tab_1}), but the RV of UBC 479 (39$\pm$15 km/s) comes from a single star, and an
RV of 22 km/s has also been reported \citep{Kounkel_2020AJ....160..279K}. Thus, we can safely say that at least five of the
six members most likely have RVs not far from 20 km/s. Less probable candidate members would be CWNU 1458, HXHWL 15, OC 0464, and Theia 1884 \citep{Hunt_2023A&A...673A.114H}.
\subsection{Vel OB1} \label{sec_3-17}

A candidate binary cluster involving NGC 2645 and Pismis 8 was first proposed by 
\citet{Subramaniam_1995A&A...302...86S}. \citet{Bica_2019AJ....157...12B} suggested the association of NGC 2645 and SAI 92 (=FSR 1436). 
\citet{Liu_2019ApJS..245...32L} identified a quartet involving Ruprecht 71, Pismis 8, LP 58, and LP 105 (LP 105 and LP 118 were
both identified as NGC 2659 by \citeauthor{Liu_2019ApJS..245...32L}). \citet{Casado_2021ARep...65..755C} reported a primordial group containing nine
candidate members: NGC 2645, NGC 2659, FSR 1436, Pismis 8, Gulliver 5, UBC 482, LP 58, and two
new OCs, Casado 28 and Casado 61. The pair formed by NGC 2659 and UBC 482 has been confirmed by
\citet{Song_2022A&A...666A..75S}.
The constraining parameters of the 20 likely members of Vel OB1 OSC according to data from 
\citet{Hunt_2023A&A...673A.114H} are summarized in Table \ref{tab_1}. 
Other robust candidates are detailed in Table \ref{tab_12}, giving a
total of 25 members. We revisited Casado 28 and Casado 61 using Gaia DR3. Casado 28 is at a
distance of 1.9 kpc and Casado 61 is 2.2 kpc away, either from photometry or from Bayesian statistics of
the star members in \citet{Bailer-Jones_2021AJ....161..147B}. The derived extinction of Casado 28 is $A_V$ = 1.44$\pm$0.08 mag. Regarding Casado 61, the estimated extinction is 3.4$\pm$0.2.\\

The projected size of this OSC is around 0.18 kpc, while the $plx$ and maximum likelihood distance lead to a
radial size of 0.4 kpc. Radial size appears overestimated at first sight. However, there is a noteworthy gap in $plx$ (from 0.44 to 0.46 mas), which could suggest two related groups \citep{Casado_2021ARep...65..755C}. 
The most distant of them would include Ruprecht 65, vdBH 54, Gulliver 5, HSC 2130, HSC 2163, FSR 1418, FSR
1424, OC 0482, and Casado 61. This group, shown below in Fig. \ref{fig_10}, would have a radial size of 0.16
kpc. The rest of members, shown above in Fig. \ref{fig_10}, would span over 0.22 kpc in heliocentric distance. These
values are closer to the mentioned projected size. The two related groups would be around 0.1 kpc apart.\\

\begin{figure}[!ht]
    \centering
    \includegraphics[width=0.5\textwidth]{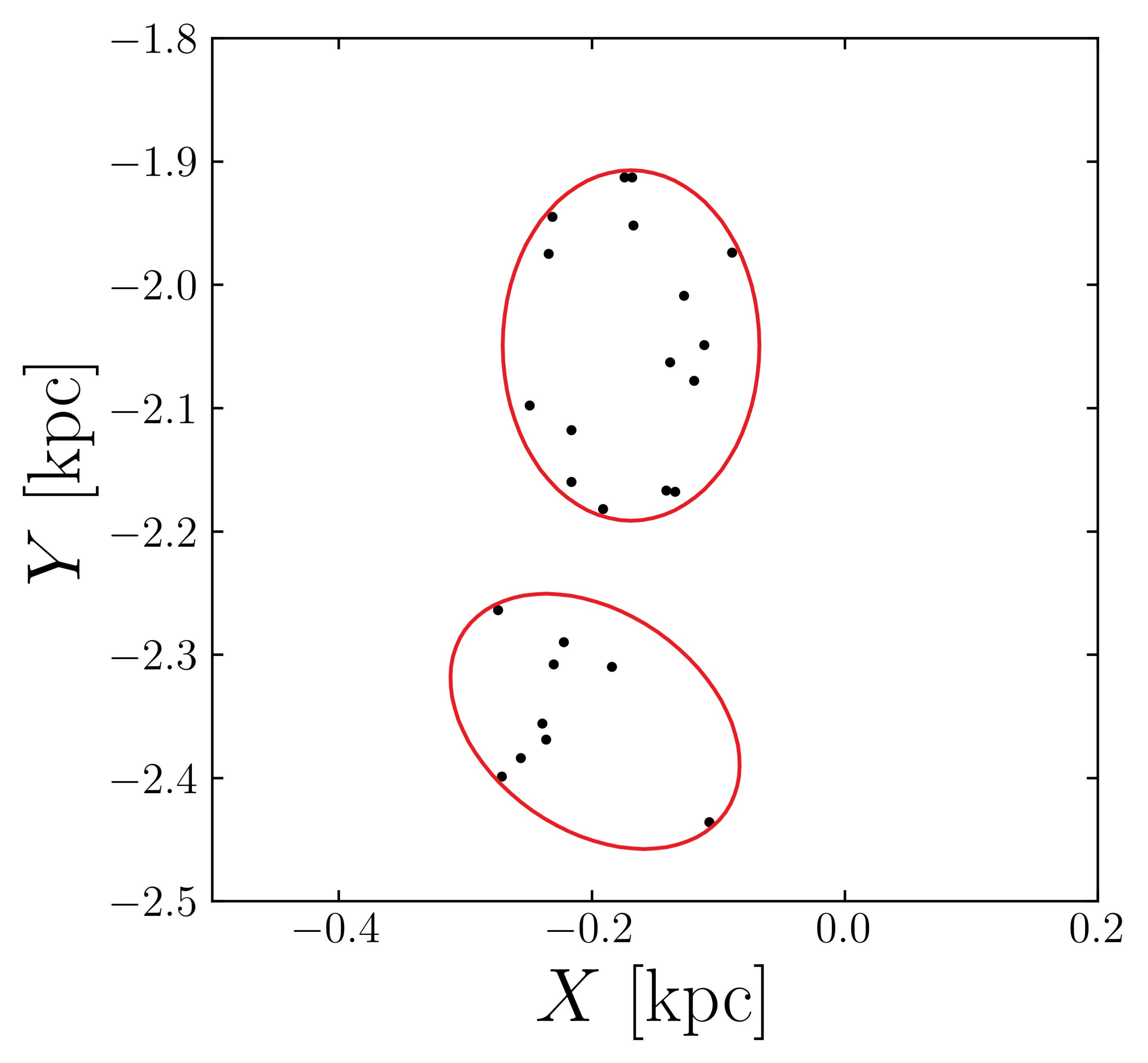}
    \caption{Positions of the cluster (sub)groups detected in Vel OB1 on the heliocentric X-Y plane of the Galaxy.}
    \label{fig_10}
\end{figure}

The tangential velocity width is 13 km/s. RVs are too dispersed to obtain a definitive conclusion, which could
be due to the existence of the two cited groups. Nevertheless, most of the reported RVs are between 17 and
43 km/s (\citealp{Casado_2021ARep...65..755C}; \citealp{Hunt_2023A&A...673A.114H}). It should be noted that HSC 2130 has an RV at odds with the
other candidate members (-172$\pm$25 km/s), but this measurement corresponds to a single star, and the rest
of its parameters match with those of its siblings. Likewise, the RV of Pismis 8
comes from a measurement of one single star (63$\pm$10 km/s; \citealp{Dias_2002A&A...389..871D}), which casts a minor doubt on its membership of this large system.
Two of the likely members of Casado 61 also have inconsistent RVs in Gaia DR3: 48$\pm$15 and -172$\pm$13
km/s. Less likely candidates could be FSR 1436, FSR 1452, Muzzio 1, UBC 245, Collinder 205, and HSC 2153
(\citealp{Liu_2019ApJS..245...32L}; \citealp{Casado_2021ARep...65..755C}; \citealp{Dias_2021MNRAS.504..356D}; \citealp{Hunt_2023A&A...673A.114H}).
\section{Discussion and conclusions} \label{sec_4}
\begin{figure}[!ht]
    \centering
    \includegraphics[width=0.5\textwidth]{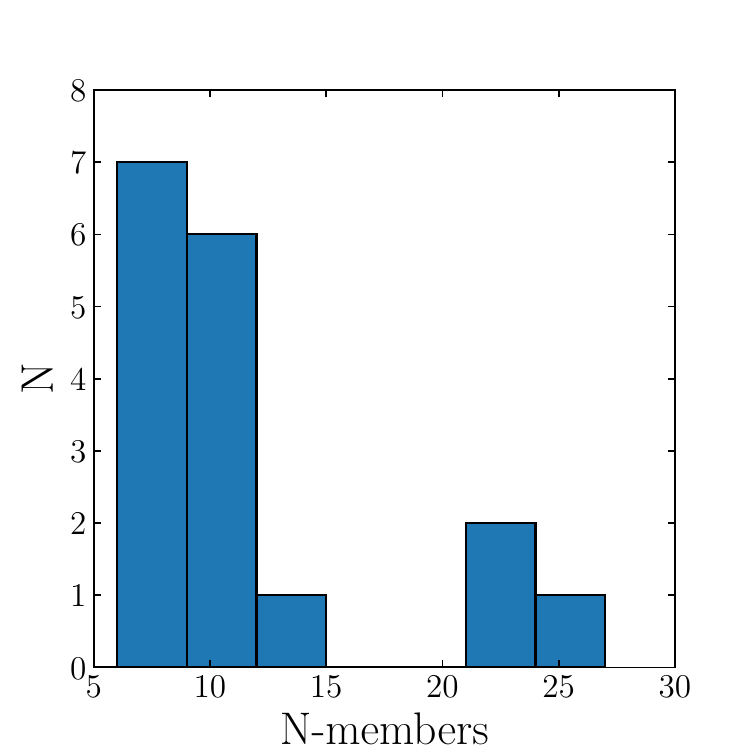}
    \caption{Histogram of the OSCs found in the third quadrant of the Galaxy as a function of their number of members.}
    \label{fig_11}
\end{figure}

We found 17 OSCs in the third quadrant of the Galactic disk, containing 190 likely
members and approximately 300 candidate members in total. Figure \ref{fig_11} displays the histogram of the OSCs
as a function of the number of members. The majority of OSCs (13 out of 17) have between 6 and 11 likely
members. It is currently unknown whether the absence of OSCs between 14 and 21 members, shown as a
gap in the histogram, is due to chance or has any significance. However, from our preliminary results
obtained for the rest of the Galaxy, it can be confidently stated that OSCs with more than 20 likely members
are rare.\\

To better understand the three anomalous OSCs having more than 20 members that we have detected,
it is noted that two of the unusual systems, Ori OB1 and Vel OB2, are located near to the Sun. Plausibly, the closer structures can be resolved into
a larger number of (sub)clusters. Conversely, Vel OB1 is not located close to us. However, in this case, at
least two distinct groups of clusters are observed in the same line of sight, each with less than 20
likely members (Fig. \ref{fig_10}). These two groups are probably related. The discovery of 17 OSCs in just one quadrant of the Galaxy at relatively short distances from the Sun suggests that OSCs are not exceptional, but rather common enough to be considered an additional class of
objects in the hierarchy of star formation. These OSCs are often associated with known OB associations
and/or star-forming regions.\\

As in previous studies ((\citealp{Casado_2021ARep...65..755C}; \citealp{Casado_2023MNRAS.521.1399C}), we identified new OCs (Casado-Hendy
2 to Casado-Hendy 5) that are members of the studied primordial groups. These findings confirm that there
are still many undiscovered OCs that can be extracted from the vast amount of Gaia data. The age of an OSC, defined as the age of its oldest member, is positively correlated with its size, as would
be expected for unbound systems, such as star associations (e.g., \citealp{Gieles_2011MNRAS.410L...6G}). The
youngest OSC in this study (CMa OB1; 14 Myr) is the smallest in size (35 x 40 pc), while the oldest OSC
(HC5; 0.25 Gyr) is one of the largest (0.20 x 0.21 kpc) and also one of the most densely populated with
stars, which helps us to understand its metastability. While the age of HC5 is considered anomalous, OSCs that
are approximately 0.15 Gyr old are not uncommon in the studied sample.\\

Many OSCs are more elongated in $l$ than in $b$ (Table \ref{tab_1}). This is not surprising, as tidal interactions that tend to disrupt the superstructure are primarily gravitational forces from the thin disk and the Galactic center. The typical scale of a GMC and OB associations (100 pc) is comparable to the size of the identified OSCs. The similar RV dispersion within the studied OSCs, of namely $\approx 15$ km/s, suggests that it is due to the uncertainty
on the mean RV, which often results from Gaia measurements of only one or a few stars in each OC. However,
for the majority of the studied OSCs, most members have RVs that are relatively close to those of their siblings.\\

It is noteworthy that the likely members of OSCs in this study were selected solely based on their matched
5D astrometric mean measurements from Gaia. However, almost all resulting members are approximately
0.1 Gyr old or younger (e.g., Table \ref{tab_1}). This provides strong evidence that only sufficiently young OCs
are close to their siblings and share similar dynamics from a common origin, while older OCs are mostly
isolated (\citealp{Casado_2022Univ....8..113C}). Most superstructures have age spans of $\ge 20$ Myr, suggesting that members of OSCs did not originate from a single burst of star formation, but rather from multiple generations within the same GMC, with new stars
forming from gas accreted from previous generations (\citealp{Li_2016Natur.529..502L}).\\

In some cases, subgroups with distinct PMs or locations are found, again indicating the
presence of multiple generations of members from several bursts of star formation. Hierarchical systems
are clearly observed in several cases, such as the Gould Belt, which encompasses several identified OSCs,
including Ori OB1, Vel OB2, and Sco-Cen. Within Ori OB1, for instance, at least two motion subgroups
have been identified, each containing several OCs. In addition, some studied clusters are disintegrating into
diverse clumps of stars (e.g., Fig. \ref{fig_6}) or subclusters (\citealp{Kuhn_2014ApJ...787..107K}). The expansion of young clusters in
associations has been observed (e.g., \citealp{Kuhn_2019ApJ...870...32K}) and is evidence that such systems are unbound. At
the bottom of the hierarchy are coronae stars of the OSC, presumably coming from dissolving clusters,
which often include known OB associations. Understanding the processes that drive this hierarchical star
formation and how it relates to the evolution of OSCs and the large-scale structure of the Galaxy is an
exciting area of research. It is clear, for instance, that OSCs are good tracers of the Galactic arms (e.g., Fig.
\ref{fig_1}). The present findings add to our understanding of the processes and structures involved in star formation
and cluster evolution, but further research is needed to fully understand the complex processes involved in
the formation and evolution of OSCs.\\

Another interesting aspect of the study of OSCs is their potential connection to the formation of GCs and
similar objects. While some works suggest that star superclusters may merge to form objects similar to GCs
(\citealp{Kroupa_1998MNRAS.300..200K}; \citealp{Fellhauer_2005ApJ...630..879F}), other evidence indicates that OSCs are typically transient and
gravitationally unbound structures. This discrepancy raises interesting questions about the formation and
evolution of GCs and their relationship to OSCs.\\

Numerical simulations by \citet{Kroupa_1998MNRAS.300..200K} proposed that stellar superclusters formed in strong starbursts
would be bound entities close to virial equilibrium. The maximum difference in RV of two OCs in the same
supercluster ($\approx 20$ km/s) seems to be in good agreement with this assumption. Under certain conditions in a
weak tidal field, these superclusters could merge after a few hundred million years to become objects similar
to faint GCs or spheroidal dwarf galaxies (\citealp{Fellhauer_2005ApJ...630..879F}). This might happen in starburst
galaxies with especially high star formation rates in the local Universe, as observed in galactic encounters,
or during the epoch when the cosmic star formation rate was at its peak (\citealp{Li_2016Natur.529..502L}; 
\citealp{Keller_2020MNRAS.495.4248K}).\\

On the other hand, our results and existing evidence (e.g., \citealp{Kuhn_2019ApJ...870...32K}) indicate that Galactic OSCs
are typically transient and gravitationally unbound structures (e.g., \citealp{Casado_2023MNRAS.521.1399C}), which tend
to disintegrate on typical timescales of 0.1 Gyr. This conclusion, along with the frequent presence of several
populations of stars with diverse metallicities in GCs and the fact that no young GCs have been observed,
suggests that the cited merging mechanism may not be working, at least in noninteracting galaxies in the
local Universe during the last several billion years.

\section*{Acknowledgements}
We thank an anonymous reviewer for insightful comments that helped to refine this study.
This work made use of data from the European Space Agency (ESA) mission Gaia
(\url{https://www.cosmos.esa.int/Gaia}, accessed on October 2022), processed by the Gaia Data Processing and
Analysis Consortium (DPAC, \url{https://www.cosmos.esa.int/web/Gaia/dpac/consortium}, accessed on
October 2022). Funding for the DPAC was provided by national institutions, in particular the institutions
participating in the Gaia Multilateral Agreement. This research made extensive use of the SIMBAD
database and the VizieR catalog access tool, operating at the CDS, Strasbourg, France (DOI:
10.26093/cds/vizier), and of NASA Astrophysics Data System Bibliographic Services. All data used in this study come from the cited astronomical databases.

\bibliographystyle{aa} 
\bibliography{references} 


\begin{appendix}
\onecolumn
\section{Catalog of OSCs from Gaia data} 
\label{appendix}

\begin{longtable}{lllllllllll}
\caption{\label{tab_1}Identified OSCs in this study.}\\
\hline\hline
1 &  2 &  3 & 4 & 5 & 6 & 7 & 8 & 9 & 10 & 11 \\
OSC & Member OCs & $\Delta l$ &$\Delta b$ & $\Delta plx$ &$\Delta d$ &$\Delta \mu^{\star}_{\alpha}$ &$\Delta \mu_{\delta}$ &  $\Delta  age$ &
$\Delta RV$ &  Ref \\
Acronym & Name &  degree & degree &mas & kpc & mas/yr & mas/yr & Myr & km/s & \\
\hline
\endfirsthead
\caption{continued.}\\
\hline\hline
1 &  2 &  3 & 4 & 5 & 6 & 7 & 8 & 9 & 10 & 11 \\
OSC & Member OCs & $\Delta l$ &$\Delta b$ & $\Delta plx$ &$\Delta d$ &$\Delta \mu^{\star}_{\alpha}$ &$\Delta \mu_{\delta}$ &  $\Delta  age$ &
$\Delta RV$ &  Ref \\
Acronym & Name &  degree & degree &mas & kpc & mas/yr & mas/yr & Myr & km/s & \\
\hline
\endhead

\endfoot
Gem OB1& HSC 1502   &188&0.5&0.49&1.71&-0.4&-1.6&6 &-8$^c$ &HR23 \\
       &  HSC 1504  &191&3.7&0.55&1.85&0.5 &-2.4&81&36     &                     \\
       &  NGC 2175  &   &   &    &    &    &    &  &       &                     \\
       &  OC 0310   &   &   &    &    &    &    &  &       &                     \\
       &  OC 0316   &   &   &    &    &    &    &  &       &                     \\
       &  Pismis 27 &   &   &    &    &    &    &  &       &                     \\
       &  UBC 1305  &   &   &    &    &    &    &  &       &                     \\
\hline
Ori OB1& ASCC 16    &199& -16&2.2&0.34&-0.1&-1.3&8 &       &CG20 \\
       & ASCC 19    &213& -20&2.9&0.41&1.6 &0.6 &98&       &                     \\
       & ASCC 21    &   &    &   &    &    &    &  &       &                     \\
       & Gulliver 6 &   &    &   &    &    &    &  &       &                     \\
       & L 1641S    &   &    &   &    &    &    &  &       &                     \\
       & NGC 1977   &   &    &   &    &    &    &  &       &                     \\
       & (=UBC 621) &   &    &   &    &    &    &  &       &                     \\
       & NGC 1981   &   &    &   &    &    &    &  &       &                     \\
       & (=UBC 207) &   &    &   &    &    &    &  &       &                     \\
       & NGC 1980   &   &    &   &    &    &    &  &       &                     \\
       & Sigma Ori  &   &    &   &    &    &    &  &       &                     \\
       & UBC 17a    &   &    &   &    &    &    &  &       &                     \\
       & UBC 17b    &   &    &   &    &    &    &  &       &                     \\
\hline
Mon OB2&    CWNU 1281    &    205&    -2.1&     0.61&    1.40&    -1.2&   -0.3&    4 &   7 &HR23 \\
       &    Collinder 104&    208&    0.2 &     0.67&    1.51&    -2.5&   1.0 &    90&   71&                     \\
       &    Collinder 106&       &        &         &        &        &       &      &     &                     \\
       &    Collinder 107&       &        &         &        &        &       &      &     &                     \\
       &    Desvoivres 2 &       &        &         &        &        &       &      &     &                     \\
       &    HSC 1629     &       &        &         &        &        &       &      &     &                     \\
       &    HSC 1636     &       &        &         &        &        &       &      &     &                     \\
       &    HSC 1641     &       &        &         &        &        &       &      &     &                     \\
       &    NGC 2244     &       &        &         &        &        &       &      &     &                     \\
\hline
Mon R2&    BDSB 91    &         213&      -10&     1.04&     0.78&     -2.1&     0.4&      4  &   23   &HR23 \\
      &    FSR 1115   &         220&      -16&     1.23&     0.86&     -3.8&     1.6&      17 &   86   &                     \\
      &    FSR 1117   &            &         &         &         &         &        &         &        &                     \\
      &    HSC 1698   &            &         &         &         &         &        &         &        &                     \\
      &    NGC 2183   &            &         &         &         &         &        &         &        &                     \\
      &    OC 0357    &            &         &         &         &         &        &         &        &                     \\
      &    Theia 15   &            &         &         &         &         &        &         &        &                     \\
      &    (= OC 0362)&            &         &         &         &         &        &         &        &                     \\
\hline
Mon OB3&   CWNU 302&       216&       -0.7&      0.41&     2.0&    -1.2&   0.5&    10&    32&HR23 \\
       &   HSC 1721&       219&       0.9 &      0.45&     2.1&    -1.9&   1.1&    96&    62&                     \\
       &   HSC 1722&          &           &          &        &        &      &      &      &                     \\
       &   HSC 1723&          &           &          &        &        &      &      &      &                     \\
       &   NGC 2311&          &           &          &        &        &      &      &      &                     \\
       &   Wit 2   &          &           &          &        &        &      &      &      &                     \\
\hline          
CMa OB1&  BDSB 96   &      224&     -0.9&      0.84&    1.11&    -3.1&    0.8&    4 &     16&HR23 \\
       &  FSR 1178  &      226&     -2.6&      0.88&    1.15&    -4.4&    1.7&    14&     33&                     \\
       &  HSC 1767  &         &         &          &        &        &       &      &       &                     \\
       &  OC 0377   &         &         &          &        &        &       &      &       &                     \\
       &  (=CMa08)  &         &         &          &        &        &       &      &       &                     \\
       &  vdBergh 92&         &         &          &        &        &       &      &       &                     \\
       &  (=CMa06)  &         &         &          &        &        &       &      &       &                     \\
\hline
HC2&      CWNU 464&      233&    -0.8&     0.34&     2.33&   -1.7&   1.9&   60 &   &HR23 \\
   &      HSC 1854&      238&    1.6 &     0.39&     2.62&   -2.2&   2.5&   160&   &                     \\
   &      HSC 1857&         &        &         &         &       &      &      &   &                     \\
   &      HSC 1893&         &        &         &         &       &      &      &   &                     \\
   &      OC 0398 &         &        &         &         &       &      &      &   &                     \\
   &      UBC 1374&         &        &         &         &       &      &      &   &                     \\
\hline
HC3&      HSC 1858  &   234&     -5&     0.33&    2.62&   -2.2&    3.1&    20&   26    &HR23 \\
   &      HSC 1859  &   238&     1 &     0.36&    3.09&   -2.6&    3.5&    90&   $59^c$&                     \\
   &      HSC 1864  &      &       &         &        &       &       &      &         &                     \\
   &      HSC 1875  &      &       &         &        &       &       &      &         &                     \\
   &      NGC 2384  &      &       &         &        &       &       &      &         &                     \\
   &      (=UBC 224)&      &       &         &        &       &       &      &         &                     \\
   &      OC 0403   &      &       &         &        &       &       &      &         &                     \\
   &      Teutsch 61&      &       &         &        &       &       &      &         &                     \\
   &      Theia 3849&      &       &         &        &       &       &      &         &                     \\
\hline             
NGC 2362&  CWNU 1064  &  238&    -4.2&    0.73&    1.17&   -2.0&  1.8&   5       &  32&     HR23 \\
        &  Ivanov 6   &  241&    -7.1&    0.81&    1.31&   -4.3&  4.3&   $1096^c$&  42&                          \\
        &  Majaess 88 &     &        &        &        &       &     &           &    &                          \\
        &  (=FSR 1297)&     &        &        &        &       &     &           &    &                          \\
        &  NGC 2354   &     &        &        &        &       &     &           &    &                          \\
        &  NGC 2362   &     &        &        &        &       &     &           &    &                          \\
        &  Theia 2267 &     &        &        &        &       &     &           &    &                          \\
\hline
HC4&     NGC 2453 &    242&      -1.0&     0.20&     4.4&     -2.3&    3.1&       25&       $^c$&    C21\\
   &     FSR 1315 &    245&      0.6 &     0.21&     4.6&     -2.6&    3.4&       81&        &               \\
   &     Casado 34&       &          &         &        &         &       &         &        &               \\
   &     Casado 35&       &          &         &        &         &       &         &        &               \\
   &     Casado 36&       &          &         &        &         &       &         &        &               \\
   &     Casado 55&       &          &         &        &         &       &         &        &               \\
\hline
Vel OB2&    vdBH 23      &244.9&   -17&   2.2&    0.29&    -4.6&    4.3&    5 &    -5&HR23 \\
       &    Collinder 135&273.3&   3  &   3.4&    0.43&    -13 &    11 &    37&    31&                     \\
       &    Collinder 140&     &      &      &        &        &       &      &      &                     \\
       &    CWNU 1020    &     &      &      &        &        &       &      &      &                     \\
       &    CWNU 1044    &     &      &      &        &        &       &      &      &                     \\
       &    CWNU 1069    &     &      &      &        &        &       &      &      &                     \\
       &    CWNU 1083    &     &      &      &        &        &       &      &      &                     \\
       &    CWNU 1096    &     &      &      &        &        &       &      &      &                     \\
       &    HSC 2056     &     &      &      &        &        &       &      &      &                     \\
       &    NGC 2451B    &     &      &      &        &        &       &      &      &                     \\
       &    NGC 2547     &     &      &      &        &        &       &      &      &                     \\
       &    OC 0450      &     &      &      &        &        &       &      &      &                     \\
       &    OC 0470      &     &      &      &        &        &       &      &      &                     \\
       &    OC 0479      &     &      &      &        &        &       &      &      &                     \\
       &    OCSN 82      &     &      &      &        &        &       &      &      &                     \\
       &    Pozzo 1      &     &      &      &        &        &       &      &      &                     \\
       &    Theia 31     &     &      &      &        &        &       &      &      &                     \\
       &    Trumpler 10  &     &      &      &        &        &       &      &      &                     \\
       &    UPK 540      &     &      &      &        &        &       &      &      &                     \\
\hline
HC5&   FSR 1342   &       246&     -1.3&       0.22&       2.7&      -2.3&   2.6&    15 &   64&        D21\\
   &   LP 201     &       250&     1.7 &       0.25&       3.8&      -2.6&   3.2&    257&   79&                  \\
   &   LP 204     &          &         &           &          &          &      &       &     &                  \\
   &   LP 942     &          &         &           &          &          &      &       &     &                  \\
   &   LP 967     &          &         &           &          &          &      &       &     &                  \\
   &   LP 1457    &          &         &           &          &          &      &       &     &                  \\
   &   Ruprecht 47&          &         &           &          &          &      &       &     &                  \\
   &   Ruprecht 48&          &         &           &          &          &      &       &     &                  \\
\hline
HC6&    NGC 2439    &   246&    -3.2&     0.23&    &-2.0&   3.1&   6  &    $51^c$&       C21\\
   &    Ruprecht 35 &   250&    -5.5&     0.26&    &-2.3&   3.3&   158&    142   &                  \\
   &    Haffner 15  &      &        &         &    &    &      &      &          &                  \\
   &    Bochum 15   &      &        &         &    &    &      &      &          &                  \\
   &    Casado 56   &      &        &         &    &    &      &      &          &                  \\
   &    UBC 238     &      &        &         &    &    &      &      &          &                  \\
   &    FSR 1351    &      &        &         &    &    &      &      &          &                  \\
   &    FSR 1352    &      &        &         &    &    &      &      &          &                  \\
\hline
Sco-Cen&   HSC 2156 &    252&     -8&      5.1&    0.132&    -16&   10&    19&   13&   HR23 \\
       &   IC 2391  &    290&      4&      7.5&    0.194&    -25&   24&    36&   25&                        \\
       &   IC 2602  &       &       &         &         &       &     &      &     &                        \\
       &   NGC 2451A&       &       &         &         &       &     &      &     &                        \\
       &   Platais 8&       &       &         &         &       &     &      &     &                        \\
       &   Platais 9&       &       &         &         &       &     &      &     &                        \\
\hline
Vela     &  vdBH 56       &    254&    -2.0&     1.02&    &-3.7&    3.4&    5  &    &LP19\\
molecular&  Alessi 43     &    265&     1.7&     1.08&    &-6.0&    5.3&    98 &    &                \\
ridge    &  Collinder 197 &       &        &         &    &    &       &       &    &                \\
         &  MWSC 1508     &       &        &         &    &    &       &       &    &                \\
         &  NGC 2546      &       &        &         &    &    &       &       &    &                \\
         &  Ruprecht 64   &       &        &         &    &    &       &       &    &                \\
\hline
HC7&      HSC 2091      &259.5&    -2.4&    0.65&     1.35&     -4.3&   4.4&    6   &  11     &  HR23 \\
   &      Majaess 99    &260.9&    -4.5&    0.70&     1.45&     -5.8&   5.3&    154 &  $39^c$ &                       \\
   &      OC 0471       &     &        &        &         &         &      &        &         &                       \\
   &      PHOC 1        &     &        &        &         &         &      &        &         &                       \\
   &      UBC 479       &     &        &        &         &         &      &        &         &                       \\
   &      (=Theia 1079) &     &        &        &         &         &      &        &         &                       \\
   &       UBC 1584     &     &        &        &         &         &      &        &         &                       \\
\hline
Vel OB1&    VdBH 54     &  263&       -2.6&        0.41&      1.9&     -4.7&    4.2&     4  &    -172&   HR23 \\
       &    CWNU 2907   &  268&        0.5&        0.51&      2.4&     -5.9&    5.4&     154&     47 &                        \\
       &    FSR 1418    &     &           &            &         &         &       &        &        &                        \\
       &    FSR 1424    &     &           &            &         &         &       &        &        &                        \\
       &    (=UBC 1444) &     &           &            &         &         &       &        &        &                        \\
       &    Gulliver 5  &     &           &            &         &         &       &        &        &                        \\
       &    HSC 2118    &     &           &            &         &         &       &        &        &                        \\
       &    HSC 2130    &     &           &            &         &         &       &        &        &                        \\
       &    HSC 2149    &     &           &            &         &         &       &        &        &                        \\
       &    HSC 2155    &     &           &            &         &         &       &        &        &                        \\
       &    HSC 2163    &     &           &            &         &         &       &        &        &                        \\
       &    NGC 2659    &     &           &            &         &         &       &        &        &                        \\
       &    OC 0480     &     &           &            &         &         &       &        &        &                        \\
       &    Ruprecht 65 &     &           &            &         &         &       &        &        &                        \\
       &    Ruprecht 71 &     &           &            &         &         &       &        &        &                        \\
       &    UBC 482     &     &           &            &         &         &       &        &        &                        \\
       &    UBC 483     &     &           &            &         &         &       &        &        &                        \\
       &    UBC 1447    &     &           &            &         &         &       &        &        &                        \\
       &    UBC 1449    &     &           &            &         &         &       &        &        &                        \\
       &    UBC 1451    &     &           &            &         &         &       &        &        &                        \\
       &    UBC 1585    &     &           &            &         &         &       &        &        &                        \\
\hline  
\end{longtable}

\tablefoot{OSCs ordered by increasing minimum Galactic longitude, their likely members, and the ranges of parameters of such members, from Gaia-derived data in the cited references. Column headings: 1. OSC acronym; 2. Members names of OCs in OSC; 3. Range in Galactic longitude; 4. Range in Galactic latitude; 5. Range in parallax; 6. Range in photometric distance; 7. Range in $\mu^{\star}_{\alpha}$; 8. Range in $\mu_{\delta}$; 9. Range in age; 10. Range in radial velocity; 11. References (HR23 \citep{Hunt_2023A&A...673A.114H}; CG20 \citep{Cantat-Gaudin_2020A&A...640A...1C}; C21 \citep{Casado_2021ARep...65..755C};
D21 \citep{Dias_2021MNRAS.504..356D}; LP19 \citep{Liu_2019ApJS..245...32L}). Abbreviations: $^c$ see text. The ranges in columns 3 to 10 refer to spans in the constraining parameters. The two rows with numbers refer to the minimum and maximum values of the cited ranges ($\Delta l$, $\Delta b$, $\Delta plx$, $\Delta d$, $\Delta \mu^{\star}_{\alpha}$, $\Delta \mu_{\delta}$, $\Delta  age$, and $\Delta RV$) for the whole OSC, not to the values for individual members. 
}

\newpage
\begin{table*}
\caption{\centering{Additional likely members of the OSC Gem OB1.}}
\label{tab_2}
\small
\begin{tabular}{lllllllllllll}
\hline\hline
1  &   2                &  3       & 4       &   5       &   6     & 7                & 8                & 9        &  10    &  11   &  12      &   13                           \\ 
OSC &   OC               &  $l$       & $b$       &   $plx$     &   d     & $\mu^{\star}_{\alpha}$   & $\mu_{\delta}$   & R        &  $N$     &  Age  &  RV      &   Ref                 \\ 
 Acronym  &   Name             &  degree  & degree  &   mas    &   kpc   & mas/yr   & mas/yr    & arcmin   &  stars &  Myr  &  km/s    &        \\ 
\hline
Gem OB1  &   NGC 2129 &    186.59  &  0.15 &    0.49  &  1.81 &   0.29  &  -2.25 &  3$^a$ & 79 & 17 &   &   D21  \\
         &   UBC 437  &    187.36  &  -1.07&    0.47  &  1.84 &   0.15  &  -1.92 &8.4$^a$ & 86 & 79 &   &   D21           \\
         &   CWNU 39  &    187.64  &   2.04&    0.51  &       &  -0.32  &  -1.76 &        & 67 & 32 &   &   H22    \\
\hline
\end{tabular}
\tablefoot{Column headings: 1. OSC acronym; 2. OC name; 3. Galactic longitude; 4. Galactic latitude; 5. Parallax; 6. Photometric distance; 7. PM in right ascension; 8. PM in declination; 9. OC radius; 10. Number of member stars; 11. Age; 12. Radial velocity. 13. References (D21 \citep{Dias_2021MNRAS.504..356D}; H22 \citep{He_2022ApJS..260....8H}). Abbreviations: $^a$ radius containing 50\% of members. 
 }
\end{table*}

\begin{table*}
\caption{\centering{Ori OB1 extra candidate member properties.}}
\label{tab_3}
\small
\begin{tabular}{lllllllllllll}
\hline\hline
1  &   2                &  3       & 4       &   5       &   6     & 7                & 8                & 9        &  10    &  11   &  12      &   13                           \\ 
OSC &   OC               &  $l$       & $b$       &   $plx$     &   d     & $\mu^{\star}_{\alpha}$   & $\mu_{\delta}$   & R        &  $N$     &  Age  &  RV      &   Ref                 \\ 
 Acronym  &   Name             &  degree  & degree  &   mas       &   kpc   & mas/yr   & mas/yr    & arcmin   &  stars &  Myr  &  km/s    &            \\ 
\hline
Ori OB1   & Collinder 70 &  205.79&    -17.24 &    2.66&           &    1.65  &    -0.94 &    $204^b$ &   493  &  4      &      &  LP19       \\
          &  (=LP 2370)  &        &           &        &           &          &          &            &        &         &      &                        \\ 
          &  ASCC 20     &  201.29&    -17.54 &    2.68&           &    -0.29 &    0.52  &    $86^b$  &   98   &  19     &      &  LP19                      \\ 
          &  (=LP 2371)  &        &           &        &           &          &          &            &        &         &      &                        \\
          &  NGC 1976    &  209.17&    -19.45 &    2.55&           &    1.15  &    0.05  &    $172^b$ &   2490 &  9      &      &  LP19                      \\
          &  (=LP 2373)  &        &           &        &           &          &          &            &        &         &      &                        \\
          &  LP 2376     &  202.70&     -19.18&    2.85&           &    1.34  &    -0.55 &    $111^b$  &   63   &  19     &      & LP19                       \\
          &  ASCC 18     &  202.39&    -18.89 &    2.42&      0.40 &    0.16  &    1.37  &    $18^a$  &   82   &  9      &  26  &  HR23   \\
          &  OC 0339     &  204.03&    -17.69 &    2.76&      0.35 &    1.67  &    -1.03 &    $16^a$  &   19   &  $79^c$ &  29  &  HR23                      \\
          &  HSC 1633    &  206.75&    -21.73 &    2.63&      0.37 &    1.84  &    -1.04 &    $33^a$  &   68   &  6      &  15  &  HR23                      \\
          &  NGC 2068    &  205.26&    -14.31 &    2.42&      0.39 &    -0.44 &    -0.73 &    $11^a$  &   102  &  6      &  6   &  HR23                      \\
          &  OCSN 59     &  203.57&    -24.60 &    2.58&      0.38 &    1.20  &    -0.86 &    $34^a$  &   60   &  6      &  17  &  HR23                      \\
          &  NGC 2024    &  206.51&    -16.37 &    2.53&           &    0.14  &    -0.75 &    $9^a$   &   62   &  4      &  27  &  Q23             \\
          &  (=OCSN 66)  &        &           &        &           &          &          &            &        &         &      &                        \\
          &  Trapezium   &  209.04&    -19.38 &    2.56&      0.38 &    1.26  &     0.27 &     $21^a$ &    269 &   6     &   24 &  D21           \\
          &  Morgan 8    &  210.11&    -19.61 &    2.63&           &    0.98  &     0.61 &            &    35  &   16    &      &  H21            \\
\hline
\end{tabular}
\tablefoot{Column headings are as in Table \ref{tab_2}. References (LP19 \citep{Liu_2019ApJS..245...32L}; HR23 \citep{Hunt_2023A&A...673A.114H}; 
Q23 \citep{Qin_2023ApJS..265...12Q}; D21 \citep{Dias_2021MNRAS.504..356D}; H21 \citep{Hao_2021A&A...652A.102H}). Abbreviations: $^a$ radius containing 50\% of members; $^b$ maximum cluster member distance to the average position; $^c$ see text. 
}
\end{table*}

\begin{table*}
\caption{\centering{Mon OB3 extra candidate member properties.}}
\label{tab_4}
\small
\begin{tabular}{lllllllllllll}
\hline\hline
1  &   2                &  3       & 4       &   5       &   6     & 7                & 8                & 9        &  10    &  11   &  12      &   13                           \\ 
OSC &   OC               &  $l$       & $b$       &   $plx$     &   d     & $\mu^{\star}_{\alpha}$   & $\mu_{\delta}$   & R        &  $N$     &  Age  &  RV      &   Ref                 \\ 
 Acronym  &   Name             &  degree  & degree  &   mas       &   kpc   & mas/yr   & mas/yr    & arcmin   &  stars &  Myr  &  km/s    &            \\ 
\hline
Mon OB3 &   BDSB 88     & 217.31  &  -1.38  &  0.49  &        &    -1.04   &    0.18   &     &    29   &     25    &      &      H21    \\
        &    Bip 14     &  217.35 &   -0.05 &   0.46 &        &     -0.97  &     0.38  &     &     34  &      22   &      &      H21    \\
        &    BDSB 90    &  218.16 &   -0.39 &   0.47 &        &     -1.29  &     0.03  &     &     16  &      24   &      &      H21    \\
        &    Ivanov 9   &  217.49 &   -0.03 &   0.50 &        &     -1.08  &     0.51  &     &     28  &       3   &      &      H21    \\
\hline
\end{tabular}
\tablefoot{Column headings are as in Table \ref{tab_2}. References (H21 \citep{Hao_2021A&A...652A.102H}). 
}
\end{table*}

\begin{table*}
\caption{\centering{CMa OB1 extra candidate member properties.}}
\label{tab_5}
\small
\begin{tabular}{lllllllllllll}
\hline\hline
1  &   2                &  3       & 4       &   5       &   6     & 7                & 8                & 9        &  10    &  11   &  12      &   13                           \\ 
OSC &   OC               &  $l$       & $b$       &   $plx$     &   d     & $\mu^{\star}_{\alpha}$   & $\mu_{\delta}$   & R        &  $N$     &  Age  &  RV      &   Ref                 \\ 
 Acronym  &   Name             &  degree  & degree  &   mas       &   kpc   & mas/yr   & mas/yr    & arcmin   &  stars &  Myr  &  km/s    &            \\ 
\hline
CMa OB1   &   Casado-Hendy 4     & 224.42  &  -2.70 &  0.87           &  1.07          &  -3.8     &  1.5         &  15  & 44  &  2        
          & 33  &  TW\\
          &                      &         &        &      $\pm$0.09  &  $\pm$0.07     &  $\pm$0.2 &  $\pm$0.3    &      &     &   $\pm$1  & $\pm$6  &           \\       
          &   Casado-Hendy 5     & 224.77  &  -1.74 &  0.83           &  1.20          &  -4.1     &  1.2         &  3   & 37  &  4        &         &  TW\\
          &                      &         &        &  $\pm$0.09      &      $\pm$0.09 &  $\pm$0.8 &   $\pm$0.6   &      &     &  $\pm$1   &         &          \\
\hline
\end{tabular}
\tablefoot{Column headings are as in Table \ref{tab_2}. References (TW (This Work)). 
}
\end{table*}

\begin{table*}
\caption{\centering{Mean parameters of the new OC Casado-Hendy 2 and UBC 1380, likely members of the OSC HC3.}}
\label{tab_6}
\small
\begin{tabular}{lllllllllllll}
\hline\hline
1  &   2                &  3       & 4       &   5       &   6     & 7                & 8                & 9        &  10    &  11   &  12      &   13                           \\ 
OSC &   OC               &  $l$       & $b$       &   $plx$     &   d     & $\mu^{\star}_{\alpha}$   & $\mu_{\delta}$   & R        &  $N$     &  Age  &  RV      &   Ref                 \\ 
 Acronym   &   Name             &  degree  & degree  &   mas       &   kpc   & mas/yr   & mas/yr    & arcmin   &  stars &  Myr  &  km/s    &            \\ 
\hline
HC3&Casado-Hendy 2  & 234.46&  -4.01 &  0.31         &  2.83         &  -2.24         &  3.14        & 19&  16&  54      &  &TW\\
   &                &       &        &      $\pm$0.04&      $\pm$0.06&       $\pm$0.04&     $\pm$0.04&   &    &  $\pm$4  &  & \\        
   &UBC 1380        & 235.04&  -3.29 &  0.35         &  2.92         &  -2.57         &  3.17        &   &  32&  11      &  &CG22\\
\hline
\end{tabular}
\tablefoot{Column headings are as in Table \ref{tab_2}. References (TW (This Work); CG22 \citep{Castro-Ginard_2022A&A...661A.118C}). 
}
\end{table*}

\begin{table*}
\caption{\centering{HC4 extra candidate member properties.}}
\label{tab_7}
\small
\begin{tabular}{lllllllllllll}
\hline\hline
1  &   2                &  3       & 4       &   5       &   6     & 7                & 8                & 9        &  10    &  11   &  12      &   13                           \\ 
OSC &   OC               &  $l$       & $b$       &   $plx$     &   d     & $\mu^{\star}_{\alpha}$   & $\mu_{\delta}$   & R        &  $N$     &  Age  &  RV      &   Ref                 \\ 
 Acronym  &   Name             &  degree  & degree  &   mas       &   kpc   & mas/yr   & mas/yr    & arcmin   &  stars &  Myr  &  km/s    &            \\ 
\hline
HC4&    UBC 640     &   245.54 &  1.51   &  0.185    &   4.7   &  -2.5   &   3.2   &  $6.1^a$  &   39  &  87 &          &   CG20    \\
   &    Haffner 18  &   243.16 &  0.45   &  0.185    &   5.1   &  -2.5   &   2.7   &  $1.9^a$  &   62  &  14 &  $60^c$  &   CG20     \\
   &    UBC 465     &   242.65 &  1.46   &  0.187    &   4.8   &  -2.4   &   2.7   &  $9.7^a$  &   78  &  41 &          &   CG20    \\
   &    Waterloo 3  &   242.56 &  1.44   &  0.19     &   4.8   &  -2.4   &   2.7   &           &       &  40 &  $65^c$  &   C21      \\
   &    Casado 57   &   245.21 &  1.60   &  0.20       &   4.6     &  -2.5    &   3.2     &  9       &   42  &  72 &     &   TW       \\
   &                &          &         &  $\pm$0.04  &   $\pm$0.2&  $\pm$0.2&   $\pm$0.2&          &       &  $\pm$5   &   &       \\   
   &    HSC 1945    &   242.64 &  -0.22  &  0.21     &   4.3   &  -2.3   &   3.0   &  $4.3^a$  &   22  &  76 &          &   HR23     \\
   &    HSC 1967    &   244.18 &  -0.33  &  0.205    &   4.5   &  -2.4   &   3.2   &  $3.2^a$  &   22  &  83 &          &   HR23      \\
   &    UBC 1400    &   244.83 &  -0.39  &  0.212    &   4.4   &  -2.4   &   3.0   &  $3^a$    &   36  &  83 &          &   HR23      \\
\hline
\end{tabular}
\tablefoot{Column headings are as in Table \ref{tab_2}. References (CG20 \citep{Cantat-Gaudin_2020A&A...640A...1C}; C21 \citep{Casado_2021ARep...65..755C}; 
TW (This Work); HR23 \citep{Hunt_2023A&A...673A.114H}). Abbreviations: $^a$ radius containing 50\% of members; $^c$ see text. 
}
\end{table*}

\begin{table*}
\caption{\centering{Vel OB2 extra candidate member properties.}}
\label{tab_8}
\small
\begin{tabular}{lllllllllllll}
\hline\hline
1  &   2                &  3       & 4       &   5       &   6     & 7                & 8                & 9        &  10    &  11   &  12      &   13                           \\ 
OSC &   OC               &  $l$       & $b$       &   $plx$     &   d     & $\mu^{\star}_{\alpha}$   & $\mu_{\delta}$   & R        &  $N$     &  Age  &  RV      &   Ref                 \\ 
 Acronym  &   Name             &  degree  & degree  &   mas       &   kpc   & mas/yr   & mas/yr    & arcmin   &  stars &  Myr  &  km/s    &            \\ 
\hline
Vel OB2  & LP 2396          &  263.37 &  -10.14 &  2.44 &         &   -5.34 &  8.21 &  81  &  90  &  13 &   &  LP19      \\
         & Alessi-Teutsch 6 &  260.41 &  -10.33 &  2.51 &         &   -4.72 &  8.96 &  115 &  244 &  7  &   &  LP19      \\
         & UBC 7            &  248.74 &  -13.35 &  3.56 &   0.276 &   -9.70 &  6.98 &  $36^a$ &  77  &  32 &   &  CG20   \\
\hline
\end{tabular}
\tablefoot{Column headings are as in Table \ref{tab_2}. References (LP19 \citep{Liu_2019ApJS..245...32L}; CG20 \citep{Cantat-Gaudin_2020A&A...640A...1C}). Abbreviations: $^a$ radius containing 50\% of members. 
}
\end{table*}

\begin{table*}
\caption{\centering{HC5 extra candidate member properties.}}
\label{tab_9}
\small
\begin{tabular}{lllllllllllll}
\hline\hline
1  &   2                &  3       & 4       &   5       &   6     & 7                & 8                & 9        &  10    &  11   &  12      &   13                           \\ 
OSC &   OC               &  $l$       & $b$       &   $plx$     &   d     & $\mu^{\star}_{\alpha}$   & $\mu_{\delta}$   & R        &  $N$     &  Age  &  RV      &   Ref                 \\ 
 Acronym  &   Name             &  degree  & degree  &   mas       &   kpc   & mas/yr   & mas/yr    & arcmin   &  stars &  Myr  &  km/s    &            \\ 
\hline
HC5   & ESO 430-18 & 248.53 &  0.78  &  0.244  &        & -2.35 &  2.75 &  15      &  65  &  91   &               &  LP19  \\
      &  HSC 2006  & 249.10 &  0.16  &  0.251  &  3.44  & -2.45 &  2.82 &  $3.9^a$ &  45  &  123  &               &  HR23   \\
      &  UBC 1420  & 248.73 &  -1.48 &  0.252  &  3.38  & -2.56 &  2.69 &  $3.6^a$ &  43  &  250  &  56           &  HR23   \\
      &            &        &        &         &        &       &       &          &      &       &  $\pm$11    &          \\
\hline
\end{tabular}
\tablefoot{Column headings are as in Table \ref{tab_2}. References (LP19 \citep{Liu_2019ApJS..245...32L}; HR23 \citep{Hunt_2023A&A...673A.114H}). Abbreviations: $^a$ radius containing 50\% of members. 
}
\end{table*}

\begin{table*}
\caption{\centering{HC6 extra candidate member properties.}}
\label{tab_10}
\small
\begin{tabular}{lllllllllllll}
\hline\hline
1  &   2                &  3       & 4       &   5       &   6     & 7                & 8                & 9        &  10    &  11   &  12      &   13                           \\ 
OSC &   OC               &  $l$       & $b$       &   $plx$     &   d     & $\mu^{\star}_{\alpha}$   & $\mu_{\delta}$   & R        &  $N$     &  Age  &  RV      &   Ref                 \\ 
 Acronym  &   Name             &  degree  & degree  &   mas       &   kpc   & mas/yr   & mas/yr    & arcmin   &  stars &  Myr  &  km/s    &            \\ 
\hline
HC6   &  UBC 469 & 247.49 &  -3.71 &  0.249  & 3.3  & -2.2 &  3.4 &  $8.3^a$ &  30  &  66  &    &  CG20      \\
      &  UBC 641 & 246.29 &  -3.59 &  0.248  &      & -2.3 &  3.2 &  $6.4^a$ &  12  &      &    &  CG20      \\
      &  FSR 1347& 248.97 &  -4.12 &  0.264  & 3.4  & -2.1 &  3.1 &  $5.2^a$ &  135 &  11  &    &  HR23      \\
\hline
\end{tabular}
\tablefoot{Column headings are as in Table \ref{tab_2}. References (CG20 \citep{Cantat-Gaudin_2020A&A...640A...1C}; HR23 \citep{Hunt_2023A&A...673A.114H}). Abbreviations: $^a$ radius containing 50\% of members. 
}
\end{table*}

\begin{table*}
\caption{\centering{Vela molecular ridge extra candidate member properties.}}
\label{tab_11}
\small
\begin{tabular}{lllllllllllll}
\hline\hline
1  &   2                &  3       & 4       &   5       &   6     & 7                & 8                & 9        &  10    &  11   &  12      &   13                           \\ 
OSC &   OC               &  $l$       & $b$       &   $plx$     &   d     & $\mu^{\star}_{\alpha}$   & $\mu_{\delta}$   & R        &  $N$     &  Age  &  RV      &   Ref                 \\ 
 Acronym  &   Name             &  degree  & degree  &   mas       &   kpc   & mas/yr   & mas/yr    & arcmin   &  stars &  Myr  &  km/s    &            \\ 
\hline
Vela        &  Pismis 5      & 259.36 &  0.91 &  1.03     & 0.94     & -5.5    &  4.3    & $7.4^a$  &  80  &  10    &         &  CG20       \\
molecular                  &           &   &   &   &         &    &   &    &    &   &   &       \\ 
 ridge                     &           &   &   &   &         &    &   &    &    &   &   &       \\ 
            &  Casado-Hendy 3  & 266.97 &  1.64 &  1.01       & 0.95       & -5.7     &  4.9       & $43^b$   &  30  &  20  & 20    &  TW   \\                      
            &                  &        &       &  $\pm$0.09  & $\pm$0.04  & $\pm$0.2 &  $\pm$0.3  &          &      &  $\pm$2    & $\pm$6   &      \\ 
\hline
\end{tabular}
\tablefoot{Column headings are as in Table \ref{tab_2}. References (CG20 \citep{Cantat-Gaudin_2020A&A...640A...1C}; TW (This Work)). Abbreviations: $^a$ radius containing 50\% of members; $^b$ maximum cluster member distance to the average position. 
}
\end{table*}

\begin{table*}
\caption{\centering{Vel OB1 supercluster extra candidate member properties.}}
\label{tab_12}
\small
\begin{tabular}{lllllllllllll}
\hline\hline
1  &   2                &  3       & 4       &   5       &   6     & 7                & 8                & 9        &  10    &  11   &  12      &   13                           \\ 
OSC &   OC               &  $l$       & $b$       &   $plx$     &   d     & $\mu^{\star}_{\alpha}$   & $\mu_{\delta}$   & R        &  $N$     &  Age  &  RV      &   Ref                 \\ 
 Acronym  &   Name             &  degree  & degree  &   mas       &   kpc   & mas/yr   & mas/yr    & arcmin   &  stars &  Myr  &  km/s    &            \\ 
\hline
Vel OB1 &      Casado 28  &   263.22 &   -2.37 &   0.50      &  1.9 &  -5.5     & 5.2      &  11  &  49  &  82       &   27         &  TW   \\
        &                 &          &         &   $\pm$0.07 &      &  $\pm$0.3 & $\pm$0.4 &      &      &  $\pm$6   &   $\pm$9  &      \\
        &      Casado 61  &   264.31 &   -1.05 &   0.43        &  2.2 &  -5.2     & 5.2      &  6  &  26 &  40       &  $^c$ &   TW \\
        &                 &          &         &   $\pm$0.09   &      &  $\pm$0.2 & $\pm$0.2 &     &     & $\pm$3    &       &    \\
        &      NGC 2645   &   264.79 &   -2.90 &   0.52    &  1.8 &  -5.9    & 5.1    &  $2.6^a$ &  36  &  30 &   17     &  C21        \\
        &      LP 58      &   264.97 &   -2.88 &   0.52    &      &  -5.8    & 5.1    &  $46^b$  &  482 &  23 &          &  C21      \\
        &      Pismis 8   &   265.11 &   -2.57 &   0.51    &      &  -5.7    & 4.9    &  $15^b$  &  64  &  47 &   63     &  C21                 \\
\hline
\end{tabular}
\tablefoot{Column headings are as in Table \ref{tab_2}. References (TW (This Work); C21 \citet{Casado_2021ARep...65..755C}). Abbreviations: $^a$ radius containing 50\% of members; $^b$ maximum cluster member distance to the average position; $^c$ see text. 
}
\end{table*}
\end{appendix}
\end{document}